\documentclass[twocolumn]{aastex701}
\usepackage{amsmath}
\shorttitle{Adaptive Gridding for SNEC}
\shortauthors{Fremling et al.}

\begin{document}

\newcommand{\Mcore}{M_{\rm core}}
\newcommand{\MNi}{M_{\rm Ni}}
\newcommand{\Msun}{{\rm M}_\odot}
\newcommand{\Rsun}{{\rm R}_\odot}
\newcommand{\Lsun}{{\rm L}_\odot}
\newcommand{\Nif}{$\rm ^{56}Ni$}

\title{SuperSNEC: Fast and Accurate Light Curve Production for Large Hydrodynamic Model Grids Using Adaptive Gridding}

\author[0000-0002-4223-103X]{Christoffer Fremling}
\affiliation{Caltech Optical Observatories, California Institute of Technology, Pasadena, CA 91125, USA}
\affiliation{Division of Physics, Mathematics and Astronomy, California Institute of Technology, Pasadena, CA 91125, USA}
\email{fremling@caltech.edu}
\author[0000-0002-0129-806X]{K-Ryan Hinds}
\affiliation{Division of Physics, Mathematics and Astronomy, California Institute of Technology, Pasadena, CA 91125, USA}
\email{khinds@caltech.edu}

\begin{abstract}
We present SuperSNEC, an accelerated version of the SuperNova Explosion Code (SNEC) designed for rapid production of large radiation-hydrodynamic model grids using low-zone-count simulations ($\sim100$ zones). The main advance is adaptive gridding of the computational grid, which preserves light-curve fidelity relative to a high-resolution SNEC baseline ($\sim1000$ zones) while delivering a runtime improvement of ${\sim}420\times$. SuperSNEC also includes solver optimizations, optimized radioactive-energy deposition and ray-tracing, improved $^{56}$Ni mixing controls, and a smooth photosphere luminosity correction that suppresses low-resolution artifacts. We quantify the speed-accuracy trade-off for a 100-zone configuration against a 1000-zone reference and define baseline settings for efficient large-grid inference of stripped-envelope supernovae. Our optimized 100-zone setup achieves an RMS light-curve residual of $0.022$~mag relative to the 1000-zone reference, at a runtime of $<2$~seconds per model. Applied to SN~2011dh (Type~IIb), SN~1993J (Type~IIb), and SN~2020oi (Type~Ic), SuperSNEC recovers light-curve parameters consistent with the literature; in particular, SN~2020oi is well reproduced by a purely radioactive model, with no clear evidence that an additional power source is required.
\end{abstract}


\section{Introduction}
The SuperNova Explosion Code (SNEC; \citealp{Morozova2015}\footnote{\url{https://stellarcollapse.org/index.php/SNEC.html}}) is widely used for one-dimensional radiation-hydrodynamics light-curve (LC) calculations of explosive transients, including a broad range of supernova (SN) applications and related fast/early-emission phenomena (e.g., \citealp{Morozova2016,Morozova2017,Piro2017,Morozova2018,Reguitti2024,Martas2025}). SNEC-based modeling has also been applied to merger-powered, kilonova-like transients (e.g., \citealp{Wu2022,Curtis2023}).

Most SNEC studies focus on modeling individual events (or small samples) using finite, low-dimensional parameter-space model grids and therefore report best-fit parameters and conditional ranges, rather than fully marginalized credible intervals that account for the full explosion-parameter space and modeling systematics. More extensive parameter-space SNEC explorations have recently been used to model Ultra-stripped SN 2023zaw \citep{Das2024} and ``superkilonova'' candidate AT~2025ulz \citep{Kasliwal2025}. However, no very large SNEC model grid ($N_{models}\sim10^6$) has yet been published for any type of transient.

The dominant driver of the SNEC runtime is the computational-grid zone count, i.e., the number of zones used within each individual model\footnote{Throughout this paper, we separate a \emph{model grid} (an ensemble of SNEC runs in parameter space) and the \emph{computational grid} within each run (the Lagrangian mass grid discretized into zones).}. A default SNEC setup with 1000 zones takes on the order of 10 minutes to run on a current CPU\footnote{Runtime $671$~seconds on an Apple silicon M1 Pro CPU. This and all other runtimes reported in this paper refer to simulations evolved to $\approx60$~d, unless otherwise noted.}. For robust unbiased light-curve parameter inference from an observed set of light curves, on the order of $1\times10^6$ models are needed to adequately map the required parameter space (Sect.~\ref{sec:grid}). This is only possible with SNEC if the number of zones is reduced to $\sim100$, which reduces the runtime to $\sim10$~seconds on the same hardware. However, with such a low-zone-count computational grid the accuracy degrades, and the degradation can manifest either at early times or at late times depending on how the zones are distributed (Sect.~\ref{sec:gridding}). When modeling SNe that exhibit any kind of early light-curve feature ($\lesssim5$~d after explosion), such as the cooling phase following shock breakout in SNe IIb (e.g., \citealp{NakarPiro2014,Piro2015}), this is unacceptable since no tradeoff between early or late-time LC accuracy can be favorably made if the goal is parameter inference via comparison to observed LCs.

Here we introduce SuperSNEC, which implements a set of performance and accuracy optimizations for SNEC, including runtime adaptive gridding of the computational grid, optimized gamma-ray deposition, improved $^{56}$Ni mixing controls, and a smooth photosphere luminosity correction. SuperSNEC reaches runtimes $<2$~s on an Apple silicon M1 Pro CPU; a practical regime in which model throughput increases by orders of magnitude while physically useful agreement with baseline SNEC (1000 zone) LCs are preserved at both very early and late phases. This makes million-model inference campaigns possible with current-generation personal computers within a few days of wall-clock time.

\section{Supernova model grids}
\label{sec:grid}

The main motivation for SuperSNEC is the combinatorial growth in the number of forward models required for robust light-curve inference when the parameter space is even moderately high-dimensional. As an example, for SNe~IIb, a minimal simulation parameter vector typically includes an extended-envelope radius and mass $(R_{\rm env},M_{\rm env})$ to capture the early shock-cooling peak, plus the bulk explosion properties (kinetic energy $E_{\rm k}$, total ejecta mass $M_{\rm ej}$) and radioactive-heating parameters (synthesized \Nif\ mass $\rm M_{Ni}$ and mixing $\rm f_{mix}$) that set the main peak and tail (e.g., \citealp{Bersten2012,MG2015}).  The early-time emission is particularly sensitive to $(R_{\rm env},M_{\rm env})$ and the explosion dynamics. The transition from the shock-cooling peak to the \Nif\ powered LC rise is particularly sensitive to $f_{\rm mix}$. 

For a precomputed model grid, the total number of models scales as the product of the number of samples per parameter,
\begin{equation}
N_{\rm models}=\prod_i N_i .
\end{equation}
Even a ``moderate'' choice of $N_i\sim6$--$10$ per dimension across six physical parameters implies $N_{\rm models}\sim(6\text{--}10)^6\approx5\times10^4$--$10^6$.  In practice, additional nuisance parameters, such as opacity-floor choices for the envelope and core used by the hydro/radiation solver, can be required to marginalize over modeling systematics. Allowing only $\sim3$--$4$ values for each of two such parameters multiplies the model-grid size by $\sim10$--$16$, pushing the total into the few$\times10^6$ regime. This motivates targeting $\sim10^6$ forward models as a practical baseline for unbiased SN~IIb light-curve inference, and it underscores why reducing per-model runtime (without sacrificing early- and late-time fidelity) is essential.
 
\section{Code Updates}
\label{sec:code_updates}

We describe each modification of SNEC~1.01 that is included in the SuperSNEC codebase below. The core hydrodynamics solver, equation of state, opacity tables, artificial viscosity, and radioactive decay physics are all unchanged from the original code \citep{Morozova2015}. Our modifications target (1)~how the computational grid evolves in time, (2)~how often expensive diagnostic and deposition routines are called, (3)~the nonlinear solver convergence thresholds, (4)~how $^{56}$Ni is placed in the ejecta and how its gamma-ray deposition is computed, and (5)~how the observed luminosity is extracted at the photosphere.

\subsection{Runtime adaptive grid mode}
\label{sec:gridding}

In the original SNEC, the computational grid is set once at the start of the simulation and never changes. Two options are available: a uniform grid in mass coordinate, or a grid read from a file (\texttt{Grid\-Pattern.dat}\footnote{Throughout this paper, names typeset in \texttt{monospace} refer to parameter keywords in the SNEC configuration file or to filenames that ship with the SNEC distribution, unless otherwise noted.}). With 1000 zones this is adequate, since there are enough zones to resolve both the surface layers, which are important during the first few days, when the photosphere sits near the stellar surface and the light curve is powered by shock-heated material cooling adiabatically, and the deep interior, which are important at later times, when the photosphere has receded inward and the light curve is powered by radioactive $^{56}$Ni decay heating diffusing outward.

With only $\sim100$ zones, this static approach forces a choice. If the grid is concentrated near the surface (as in the default \texttt{Grid\-Pattern.dat}), the early shock-cooling phase is well resolved, but the interior where the bulk of the ejecta mass and all of the $^{56}$Ni reside is sparsely sampled, and the late-time light curve suffers. Conversely, a uniform grid resolves the interior better but places too few zones near the surface for the early phase. No fixed 100-zone grid can do both well.

We address this by introducing a new adaptive gridding (moving-rezoning at fixed zone count) mode (e.g., \citealp{HartenHyman1983,DorfiDrury1987}), \texttt{adaptive\_\-runtime}, in which the code periodically redistributes the zone boundaries during the simulation. The idea is straightforward: at early times, concentrate zones near the surface where the photosphere is; as the photosphere recedes inward over the following days, gradually shift zones toward the interior. This mode does not require any user pre-defined input grid.

\subsubsection{Grid morphology}

At any given time the target grid has a piecewise structure: an inner region uniformly spaced in Lagrangian mass, followed by an outer region whose cell widths decrease geometrically toward the surface. This provides robust interior sampling while concentrating resolution where the photosphere is located.

We tie the target morphology directly to the position of the photosphere rather than to separate early- and late-time targets. For each remap we locate the photosphere (the surface at optical depth $\tau = 2/3$) in the Lagrangian mass coordinate $m_{\mathrm{photo}}$ and compute its fractional position within the domain,
\begin{equation}
q_{\mathrm{photo}} = \frac{m_{\mathrm{photo}} - m_1}{m_N - m_1},
\end{equation}
where $m_1$ and $m_N$ are the inner and outer mass boundaries. The fraction of the $N$ zones (where $N \equiv \texttt{imax}$ is the total zone count) assigned to the inner uniform region is
\begin{equation}
f_{\mathrm{inner}} = \min\!\bigl(0.60,\;\max(0.10,\;0.15 + 0.30\,q_{\mathrm{photo}})\bigr),
\end{equation}
so that the uniform interior region grows as the photosphere recedes inward.

The surface concentration of the outer region is controlled by a single parameter, \texttt{grid\_surface\_\-alpha} ($\alpha_{\mathrm{surf}}$). For the $N_{\mathrm{outer}} = (1 - f_{\mathrm{inner}})\,N$ outer zones we define the geometric ratio
\begin{equation}
r_{\mathrm{eff}} = \exp\!\left(-\frac{\alpha}{N_{\mathrm{outer}}}\right),
\label{eq:r_eff}
\end{equation}
where $\alpha = \alpha_{\mathrm{surf}} \times \min(1.0,\,\max(0.35,\, q_{\mathrm{photo}}))$. At early times the photosphere sits near the surface ($q_{\mathrm{photo}} \approx 1$) and the full $\alpha_{\mathrm{surf}}$ is applied, concentrating zones where they are needed most. As the photosphere recedes inward over the following days ($q_{\mathrm{photo}} \rightarrow 0$), the scaling factor decreases (floored at 0.35), relaxing the surface concentration and redistributing zones toward the interior. The cell width at outer zone $k$ ($k = 0, 1, \ldots, N_{\mathrm{outer}}-1$) is $\Delta m_k = \Delta m_{\mathrm{inner}} \times r_{\mathrm{eff}}^{\,k}$, where $\Delta m_{\mathrm{inner}}$ is the uniform inner zone width. To prevent numerically degenerate surface cells we additionally clamp $r_{\mathrm{eff}}$ so that the thinnest outer cell is at least \texttt{grid\_min\_cell\_\-frac} times the uniform cell width. The baseline value is $10^{-4}$, which is conservative enough that this floor is rarely active in practice; it serves as a safety net against pathological configurations at high $N$.

\subsubsection{Remap cadence}
\label{sec:remap_scheduling}
The grid is redistributed at a fixed interval $\Delta t_{\mathrm{remap}}$ (\texttt{grid\_update\_\-interval\_days}; baseline 1.0~d). At each scheduled check we rebuild the target grid from $q_{\mathrm{photo}}$. If the maximum boundary shift would be smaller than $10^{-4}$ of the total mass span, the remap is skipped to avoid unnecessary interpolation diffusion. The sensitivity to the choice of $\Delta t_{\mathrm{remap}}$ is quantified in Appendix~\ref{sec:grid_cadence_sweep}.

Once the photosphere has receded below $q_\mathrm{photo} \le q_\mathrm{stop}$ (\texttt{grid\_remap\_\-qphoto\_stop}; baseline $q_\mathrm{stop} = 0.50$), grid remapping is permanently disabled for the remainder of the simulation. Below this threshold the photosphere tracks the inner ejecta where the grid target changes slowly; continued remaps would introduce interpolation diffusion and redundant forced Ni recalculations (Sect.~\ref{sec:ni_remap_interaction}) without commensurate accuracy gain.

\subsubsection{Relaxation}

To limit the interpolation diffusion introduced by each remap, we blend the current grid toward the target using exponential relaxation \citep{BrackbillSaltzman1982}. For each interior zone boundary $i$ with mass coordinate $m_i$,
\begin{equation}
m_i^{\mathrm{new}} = (1 - \alpha_{\mathrm{relax}})\, m_i^{\mathrm{old}} + \alpha_{\mathrm{relax}}\, m_i^{\mathrm{target}},
\end{equation}
with
\begin{equation}
\alpha_{\mathrm{relax}} = 1 - \exp\!\left(-\frac{\Delta t}{t_{\mathrm{relax}}}\right),
\end{equation}
where $\Delta t$ is the elapsed time since the previous remap and $t_{\mathrm{relax}} \equiv \texttt{grid\_relax\_\-days}$ (baseline 5.0~d). For numerical stability we additionally cap $\alpha_{\mathrm{relax}}$ so that no boundary moves by more than 50\% of its local two-zone width in a single remap (a global limiter that uniformly scales $\alpha_{\mathrm{relax}}$ when any boundary would exceed this threshold). After blending we enforce strict monotonicity of the mass grid ($m_{i+1} > m_i$) with forward and backward sweeps.

To ensure convergent accuracy across zone counts, $t_{\mathrm{relax}}$ is scaled with $N$: the effective relaxation time is $0.4\,t_{\mathrm{relax}}$ for $N \le 80$, rising linearly to $1.0\,t_{\mathrm{relax}}$ at $N = 100$ and $2.0\,t_{\mathrm{relax}}$ for $N \ge 200$. At low $N$ the shorter timescale allows the grid to track the rapidly evolving photosphere; at high $N$ the longer timescale limits per-remap interpolation diffusion in the thinner surface cells.

\subsubsection{Conservative remapping}
\label{sec:remap}

When the grid is updated, all hydrodynamic variables must be mapped from the old zone boundaries to the new ones. We use different remapping strategies depending on where the variable is defined:

\begin{itemize}
\item \textit{Boundary-defined quantities} (radius, velocity): linear interpolation in mass coordinate from old to new boundaries.
\item \textit{Cell-centered quantities} (temperature, internal energy, composition, Ni heating): piecewise-linear conservative remapping. Within each old cell $i$, we reconstruct a linear profile $f(m) = \bar{f}_i + s_i\,(m - \bar{m}_i)$, where $\bar{f}_i$ is the cell-average value of the quantity, $\bar{m}_i$ is the cell-center mass coordinate, and $s_i$ is a slope limited by the minmod function to prevent oscillations and preserve monotonicity \citep{vanLeer1979,Sweby1984}. This reconstruction is then integrated analytically over each new cell to obtain the new cell-average value, following the standard Lagrangian-remap approach used in astrophysical hydrodynamics \citep{Blondin1993}. This gives second-order accuracy in the zone width $\Delta m$ in smooth regions while remaining monotonicity-preserving.
\end{itemize}

Temperature and internal energy are both remapped conservatively; thermodynamic consistency is restored by the subsequent implicit solve (which calls the EOS) on the remapped state.

\subsubsection{Composition handling}
For composition, we take advantage of the fact that the high-resolution progenitor profile ($\mathrm{few}\times1000$ zones from the stellar evolution\footnote{Any 1D stellar evolution code can provide progenitor profiles; e.g., MESA \citep{Paxton2011}, KEPLER \citep{WoosleyWeaver1995}, and GENEC \citep{Ekstrom2012}.}) is stored in memory at initialization. Rather than remapping the coarse 100-zone composition through successive interpolations, which would progressively smooth sharp interfaces, we re-map the stored high-resolution profile directly onto each new grid. This preserves the original composition structure regardless of how many grid updates have occurred. The mass-weighted conservative averaging preserves $\sum_i X_i = 1$ by construction; the code verifies closure after each mapping and aborts if it is violated beyond $10^{-8}$ after summing over all composition species. The stored high-resolution profile is also the substrate for our \Nif\ mixing prescription (Sect.~\ref{sec:ni_mixing}).

\subsubsection{Interaction with $^{56}$Ni heating}
\label{sec:ni_remap_interaction}

Each grid remap conservatively interpolates the $^{56}$Ni deposit function onto the new grid using the piecewise-linear scheme described in Sect.~\ref{sec:remap}. To ensure the spatial heating distribution is exact on the new grid, the Ni update timer is clamped to the current time after every remap, forcing a full $^{56}$Ni heating recalculation on the next hydro step. The number of Ni evaluations is therefore closely tied to the number of remaps.

When a regular adaptive-cadence Ni update (Sect.~\ref{sec:ni_cadence}) and a remap-forced update fall within a short time of each other, the code suppresses the duplicate: the time of the last Ni evaluation is tracked, and a regular-cadence call is skipped if it occurs within half the base grid update interval of the previous evaluation. This deduplication is applied in the solver rather than in the remap routine, so remap-forced updates always take priority and the grid always receives a fresh Ni calculation after a remap.

\subsubsection{Resolution independence}
\label{sec:imax_scaling}

The adaptive gridding mode must function correctly at any zone count $N = \texttt{imax}$, not only at the baseline $N=100$ that we plan to use for producing large model grids (Sect.~\ref{sec:baseline}). This is important for two reasons: (1)~it allows users to increase resolution for targeted high-fidelity runs of individual models of interest identified from a fast model grid, and (2)~it validates that the adaptive scheme is numerically well-posed rather than tuned to a specific resolution.

The exponential form of the geometric ratio (Eq.~\ref{eq:r_eff}) keeps the total surface concentration controlled by a single parameter across different zone counts. In addition, \texttt{grid\_min\_cell\_\-frac} enforces a minimum outer-zone mass width $\Delta m$ (the thinnest outer cell must be at least this fraction of the uniform inner cell mass width), and the relaxation limiter caps how far boundaries can move in a single remap. Together these choices keep the algorithm stable across a wide range of $N$.

We verified this by running the \texttt{adaptive\_\-runtime} mode at $N = 60$, $80$, $100$, $200$, $500$, and $1000$ zones with baseline parameters (Sect.~\ref{sec:baseline}) except for $n_\mathrm{quad} = 150$, matching the original SNEC default to ensure the gamma-ray deposition quadrature is converged at all zone counts (Appendix~\ref{sec:ni_quad}). Table~\ref{tab:imax_scaling} summarizes the results compared against a 1000-zone reference run with the original SNEC codebase (Sect.~\ref{sec:validation}). All six configurations complete successfully.

\begin{deluxetable}{rrrr}
\setlength{\tabcolsep}{6pt}
\tablecaption{Runtime and accuracy as a function of zone count $N$. SuperSNEC rows use the \texttt{adaptive\_runtime} grid mode and baseline settings (Sect.~\ref{sec:baseline}), with the SNEC default gamma-ray deposition quadrature order $n_\mathrm{quad} = 150$ (Sect.~\ref{sec:ni_quad}). RMS is measured against a 1000-zone reference run produced with the standard (unmodified) SNEC code and default \texttt{GridPattern.dat}. \label{tab:imax_scaling}}
\tablehead{
\colhead{\hfill$N$} & \colhead{$t$ (s)} & \colhead{$\Delta m_{5\text{--}30}$} & \colhead{$\Delta m_\mathrm{all}$}
}
\startdata
\cutinhead{SuperSNEC ($n_\mathrm{quad} = 150$)}
  60 &   1.2 & 0.057 & 0.045 \\
  80 &   1.6 & 0.041 & 0.032 \\
 100 &   2.3 & 0.033 & 0.027 \\
 200 &   6.7 & 0.026 & 0.024 \\
 500 &  25.5 & 0.022 & 0.021 \\
1000 &  52.4 & 0.016 & 0.017 \\
\cutinhead{Reference (SNEC 1000-zone)}
1000 & 671 & \nodata & \nodata \\
\enddata
\tablecomments{$\Delta m_{5\text{--}30}$ and $\Delta m_\mathrm{all}$ are RMS magnitude residuals in the 5--30\,d and $>$0.5\,d windows.}
\end{deluxetable}

The RMS residuals improve steadily with increasing $N$: $\Delta m_\mathrm{all} = 0.045$~mag at $N = 60$, $0.027$~mag at $N = 100$, and $0.017$~mag at $N = 1000$. Reducing the zone count from $N = 1000$ to $N = 100$ cuts the runtime by ${\sim}96\%$, from $54$~s to $2.3$~s (at $n_\mathrm{quad}=150$; Appendix~\ref{sec:ni_quad}). The $N = 1000$ adaptive run itself completes in $54$~s, approximately $12\times$ faster than the unmodified SNEC reference run ($671$~s). The aggregate runtime optimizations that produce this overall speedup in SuperSNEC are detailed in Sect.~\ref{sec:speed_results}. We note that complete parity with the SNEC reference even at $N = 1000$ is not expected. The $\Delta m_\mathrm{all} = 0.017$~mag residual floor arises from two differences in how the progenitor composition reaches the simulation grid. First, SuperSNEC smooths and normalizes on the high-resolution stored profile before mapping to the grid (Sects.~\ref{sec:ni_mixing} and~\ref{sec:ni_independent_smoothing}), whereas SNEC interpolates the raw profile onto the computational grid and smooths afterward. Second, SuperSNEC's grid projection uses conservative mass-weighted cell averaging, while SNEC uses point interpolation. Together, the different ordering and projection method produce slightly different initial conditions regardless of grid mode or zone count.

\subsection{Adaptive output and diagnostics cadence}
\label{sec:cadence}

In the original SNEC code, all diagnostic quantities (photosphere location, optical depths, luminosity shell, characteristic timescales, shell energies, bolometric magnitudes, and the energy conservation budget) are computed at every hydrodynamic timestep, even though none of them feed back into the integration. In SuperSNEC these diagnostics are evaluated only at scalar output times, i.e., the cadence at which scalar time series (light curves, photospheric velocity, magnitudes) are sampled, which is independent of and typically denser than the profile output cadence used for radial structure dumps. At 100 zones the computational cost of the diagnostics is negligible compared to the solver (Appendix~\ref{sec:output_cadence_speed}), so this is primarily a code-quality improvement rather than a speed optimization.

We introduce a three-tier output cadence system. Instead of a single fixed output interval, the code uses three cadence bands:
\begin{itemize}
\item \textit{Fast} (early times, $t < t_{\mathrm{mid}}$): frequent output to capture shock breakout and cooling.
\item \textit{Mid} (intermediate, $t_{\mathrm{mid}} \leq t < t_{\mathrm{late}}$): moderate cadence during the transition phase.
\item \textit{Late} ($t \geq t_{\mathrm{late}}$): sparse output during the radioactive tail.
\end{itemize}
Profile dumps and scalar (light-curve) dumps have independent cadence settings in each band (Table~\ref{tab:parameters}). The transition times $t_{\mathrm{mid}}$ and $t_{\mathrm{late}}$ are user-configurable (baseline 10 and 20~d). Each output interval is quantized to an integer number of hydrodynamic timesteps to ensure dumps occur exactly at step boundaries.

\subsubsection{Minimal output mode}
\label{sec:output_mode}

A standard SNEC run writes 61 output files per model: 29 radial-profile files (velocity, density, temperature, etc.), 9 scalar diagnostic time series, per-timestep solver timing logs, composition dumps, and initial-condition snapshots. At 100 zones with the baseline output cadence, the full output totals ${\sim}6$~MB per model. For single-model analysis this is useful, but for a $10^6$-model grid it amounts to ${\sim}6$~TB of output, the vast majority of which is not needed for light-curve fitting.

We introduce a minimal output mode (\texttt{output\_mode\,=\,1}) that writes only data files needed by downstream light-curve fitting and inference:
\begin{enumerate}
\item \texttt{lum\_observed.dat} --- bolometric luminosity vs.\ time,
\item \texttt{vel\_photo.dat} --- photospheric velocity vs.\ time,
\item \texttt{magnitudes.dat} --- broadband magnitudes (bolometric-correction method),
\item \texttt{info.dat} --- model metadata (mass, breakout time, etc.),
\item \texttt{run\_summary.dat} --- self-contained key--value summary of all run parameters and diagnostics (written once at end of run).
\end{enumerate}
This reduces the output from 61 files to 5 and the disk footprint from ${\sim}6$~MB to ${\sim}50$~KB per model (${\sim}50$~GB for a $10^6$-model grid, vs.\ ${\sim}6$~TB in full mode). All profile dumps, composition snapshots, per-timestep timing logs, conservation diagnostics, and initial-condition files are suppressed. The simulation itself is identical between modes; only file output is affected.

The default (\texttt{output\_mode\,=\,0}) retains the full SNEC output for single-model analysis. All runtimes reported in this paper use \texttt{output\_mode\,=\,0} unless otherwise noted; the additional runtime savings from \texttt{output\_mode\,=\,1} are quantified in Table~\ref{tab:cumulative_speedup}.

\subsection{Adaptive $^{56}$Ni deposition cadence}
\label{sec:ni_cadence}

The radioactive heating computation in SNEC (\texttt{nickel.F90}) is one of the most expensive per-call routines in the code. It evaluates the gamma-ray deposition function for each grid zone using a two-dimensional numerical integration over polar angle and radial coordinate ($150 \times 150 = 22{,}500$ integration points per zone), following the formalism of \citet{Swartz1995}. In the original code, this routine is called at a fixed interval \texttt{Ni\_period} (in seconds), regardless of the simulation time. The adaptive cadence described below applies to both fixed-grid (\texttt{legacy\_pattern}) and adaptive-grid (\texttt{adaptive\_runtime}) modes, though its effective impact differs between them (Sect.~\ref{sec:speed_results}). We show in Appendix~\ref{sec:ni_quad} that $n_\mathrm{quad}$ can be reduced substantially for 100-zone models without measurable accuracy loss.

SuperSNEC uses an adaptive $^{56}$Ni deposition cadence that tracks the physical timescale of the heating. The total specific heating rate from the $^{56}$Ni $\rightarrow$ $^{56}$Co $\rightarrow$ $^{56}$Fe decay chain is \citep{Arnett1982}
\begin{equation}
\dot{q}(t) = \epsilon_{\mathrm{Ni}}\, e^{-t/\tau_{\mathrm{Ni}}} + \epsilon_{\mathrm{Co}}\left(e^{-t/\tau_{\mathrm{Co}}} - e^{-t/\tau_{\mathrm{Ni}}}\right),
\end{equation}
where $\epsilon_{\mathrm{Ni}}$ and $\epsilon_{\mathrm{Co}}$ are the specific energy generation rates of $^{56}$Ni and $^{56}$Co decay, respectively, and $\tau_{\mathrm{Ni}} = 8.8$~d and $\tau_{\mathrm{Co}} = 111.3$~d are their mean lifetimes \citep{Nadyozhin1994}. We define a heating timescale $t_{\mathrm{heat}} = |\dot{q} / \ddot{q}|$, where $\ddot{q} \equiv d\dot{q}/dt$ is the time derivative of the heating rate, and set the update period to
\begin{equation}
\Delta t_{\mathrm{Ni}} = \mathrm{clip}\!\left(f_{\mathrm{Ni}}\, t_{\mathrm{heat}},\; \Delta t_{\mathrm{Ni,min}},\; \Delta t_{\mathrm{Ni,max}}\right),
\label{eq:ni_cadence}
\end{equation}
where the $\mathrm{clip}(x, a, b) = \min(\max(x, a), b)$ function clamps its argument to the interval $[a, b]$, $f_{\mathrm{Ni}}$ is a user-specified fractional-change target (baseline $0.20$), and $\Delta t_{\mathrm{Ni,min}}$, $\Delta t_{\mathrm{Ni,max}}$ are the lower and upper bounds (baseline $5 \times 10^4$~s and $5.456 \times 10^5$~s). The result is quantized to the nearest integer number of hydrodynamic timesteps.

At early times, when $^{56}$Ni is decaying rapidly ($t_{\mathrm{heat}} \approx \tau_{\mathrm{Ni}} = 8.8$~d), the baseline $f_{\mathrm{Ni}} = 0.20$ gives updates every ${\sim}2$~d. At late times, when the heating changes slowly on the $^{56}$Co timescale ($\tau_{\mathrm{Co}} = 111.3$~d), the computed interval exceeds $\Delta t_{\mathrm{Ni,max}}$ and is clamped to ${\sim}6$~d. This avoids both the waste of recomputing a nearly unchanged deposition function and the inaccuracy of updating too rarely during the early Ni-dominated phase. The speed--accuracy trade-off for the cadence parameters is quantified in Sect.~\ref{sec:speed_results} (Table~\ref{tab:ni_ca tdence_comparison}).

\subsubsection{Gamma-ray deposition optimizations}
\label{sec:ni_ray_interp}

The $^{56}$Ni heating calculation traces gamma-ray paths through the ejecta, accumulating optical depth from the gamma-ray mass absorption coefficient $\kappa_\gamma = 0.06\,Y_e$~cm$^{2}$\,g$^{-1}$, where $Y_e$ is the electron fraction per baryon. We apply two optimizations to this integral. First, the per-zone linear absorption coefficient $\kappa_\gamma \rho$ is precomputed into an array before the integration loops, eliminating redundant arithmetic in the innermost loop. Second, the radial zone lookup, which in the original SNEC performs a binary search for every integration point, is replaced with a sequential hunt that exploits the monotonic inward traversal of successive ray segments. Together, these changes reduce the per-evaluation cost of the deposition integral with no change to the computed heating rates.

We also replace the piecewise-constant (staircase) representation of $\kappa_\gamma \rho$ and the $^{56}$Ni mass fraction $X_\mathrm{Ni}$ along each ray with piecewise-linear interpolation between cell centers. At low zone counts, the staircase approximation introduces systematic error in the deposited energy, particularly at late times when the gamma-ray mean free path exceeds the zone width. Linear interpolation removes this artifact at negligible additional cost. The improvement is most pronounced at $N \sim 100$ and vanishes at high zone counts ($N \gtrsim 500$) where the two representations converge. Both optimizations are enabled by default.

\subsection{Solver optimizations}
\label{sec:solver}

The implicit hydrodynamic and radiation-hydrodynamic solvers in SNEC (\texttt{hydro.F90} and \texttt{hydro\_rad.F90}) use Newton--Raphson iteration to converge the coupled temperature--pressure--velocity state at each timestep. The radiation-hydrodynamic solver solves a strictly tridiagonal linear system at each iteration; the original code uses LAPACK's general banded solver \texttt{dgbsv}, and we replace it with the dedicated tridiagonal routine \texttt{dgtsv}, which eliminates the overhead of banded LU factorization and pivoting while producing identical results to machine precision.

Convergence of the Newton--Raphson iteration is declared when the maximum relative temperature change across all zones falls below a tolerance $\mathrm{EPSTOL}$, subject to a maximum iteration count $\mathrm{ITMAX}$. In the original SNEC, $\mathrm{EPSTOL}$ is hard-coded to $10^{-7}$ in both solvers, and $\mathrm{ITMAX}$ is likewise a compiled constant; changing any of these values requires editing the source and recompiling. SuperSNEC promotes all four settings to runtime parameters (\texttt{epstol\_hydro}, \texttt{epstol\_rad}, \texttt{itmax\_hydro}, \texttt{itmax\_rad}), read from the configuration file, enabling systematic benchmarking without recompilation.

We relax the default to $\mathrm{EPSTOL} = 10^{-4}$, a factor of 1000 relative to the original SNEC value. The maximum number of iterations ($\mathrm{ITMAX}$) is kept unchanged at 100 for the hydro solver and 300 for the radiation-hydro solver. The rationale is that a tolerance of $10^{-7}$ demands convergence far beyond the intrinsic accuracy of the underlying physics and numerics: the OPAL opacity tables \citep{Iglesias1996} are uncertain at the percent level and are interpolated in $(\rho,T)$, our Saha ionization treatment uses a simplified three-species approximation, and the artificial viscosity \citep{VonNeumann1950} smears shocks over a few zone widths. A tolerance of $10^{-4}$ is therefore tight enough that solver convergence is not the limiting source of error. We verify this empirically through an extensive threshold sweep in Appendix~\ref{sec:solver_sweep}.

\subsection{Flexible $^{56}$Ni mixing prescription}
\label{sec:ni_mixing}

In the original SNEC, radioactive $^{56}$Ni is placed into the ejecta using a step function (also referred to as a ``box'' profile). The user specifies two parameters: \texttt{Ni\_mass} (the total $^{56}$Ni mass in solar masses) and \texttt{Ni\_boundary\_\-mass} (the outer boundary of the Ni-containing region, as an absolute mass coordinate in solar masses). All zones interior to \texttt{Ni\_boundary\_\-mass} receive a uniform Ni mass fraction $X_{\mathrm{Ni}} = M_{\mathrm{Ni}} / (M_{\mathrm{boundary}} - M_{\mathrm{cut}})$, where $M_{\mathrm{Ni}}$ is the total synthesized $^{56}$Ni mass, $M_{\mathrm{boundary}}$ is the outer boundary mass coordinate, and $M_{\mathrm{cut}}$ is the mass coordinate of the inner boundary (the mass cut, set by the \texttt{mass\_excised} parameter when \texttt{mass\_excision\,=\,1}). All zones exterior to \texttt{Ni\_boundary\_\-mass} receive $X_{\mathrm{Ni}} = 0$.

This approach has three limitations for large model-grid applications. First, \texttt{Ni\_boundary\_\-mass} is an absolute mass coordinate that depends on the specific progenitor model: a value appropriate for a $3\,\Msun$ helium star is not appropriate for a $5\,\Msun$ one. This makes it difficult to define a single set of parameters that works across a range of progenitors with varying properties. Second, the step function produces a sharp discontinuity in composition at \texttt{Ni\_boundary\_\-mass}, which is unphysical. In real SN explosions, hydrodynamic instabilities partially mix $^{56}$Ni outward with a smooth falloff \citep{Kifonidis2003}. SNEC includes optional boxcar smoothing of the $^{56}$Ni profile to mitigate this, which produces reasonable profiles when the smoothing width and mixing boundary are chosen carefully. However, both the Ni placement and the boxcar smoothing operate on the simulation grid, so with only $\sim100$ zones the smoothed profile is resolution-limited.

We replace this with a prescription based on the ejecta mass fraction $q = (m - M_{\mathrm{cut}}) / M_{\mathrm{ej}}$, which ranges from 0 at the mass cut to 1 at the ejecta surface ($m$ is the Lagrangian mass coordinate and $M_{\mathrm{ej}}$ is the total ejecta mass). Instead of an absolute mass coordinate, the user specifies \texttt{Ni\_mix\_\-fraction} ($q_{\mathrm{mix}}$, a value between 0 and 1), which sets the fraction of the total ejecta mass over which $^{56}$Ni is distributed. This is progenitor-independent: $q_{\mathrm{mix}} = 0.3$ means that the innermost 30\% of the ejecta mass contains Ni, regardless of the total ejecta mass.

In addition to $q_{\mathrm{mix}}$, the user selects a kernel shape via \texttt{Ni\_mix\_\-kernel}. Kernels~1 and~2 are fully specified by $q_{\mathrm{mix}}$ alone, while Kernel~3 introduces two optional parameters for a two-component profile. The three options are:
\begin{itemize}
\item \textit{Kernel 1 (step function):} Identical physics to the original SNEC approach, but parametrized by mass fraction rather than absolute mass. $X_{\mathrm{Ni}}$ is uniform within the mixing region and zero outside. If the requested $M_{\mathrm{Ni}}$ cannot fit within $q_{\mathrm{mix}}$ (i.e., $X_{\mathrm{Ni}}$ would exceed unity), $q_{\mathrm{mix}}$ is automatically expanded until $X_{\mathrm{Ni}} \leq 0.95$; the same auto-expansion applies to both components of Kernel~3.
\item \textit{Kernel 2 (flat core + exponential tail):} The default. The inner half of the mixing region ($q \leq q_{\mathrm{mix}}/2$) has a flat (uniform) Ni profile. The outer half ($q_{\mathrm{mix}}/2 < q \leq q_{\mathrm{mix}}$) follows an exponential decay. The unnormalized Ni profile shape $\phi_{\mathrm{Ni}}$ as a function of $q$ is
\begin{equation}
\phi_{\mathrm{Ni}}(q) = \begin{cases}
1 & q \leq q_0, \\[6pt]
\exp\!\Bigl(-\dfrac{q - q_0}{\delta q}\Bigr) & q_0 < q \leq q_{\mathrm{mix}}, \\[6pt]
0 & q > q_{\mathrm{mix}},
\end{cases}
\end{equation}
where $q_0 = q_{\mathrm{mix}}/2$ is the transition point and $\delta q = q_{\mathrm{mix}}/6$ is the $e$-folding scale. This profile is then normalized so that the integrated Ni mass equals \texttt{Ni\_mass}.
\item \textit{Kernel 3 (two-component boxcar):} A two-zone uniform profile that splits the $^{56}$Ni mass between a core component and an extended outer shell. A fraction $f_{\mathrm{outer}}$ (\texttt{Ni\_mix\_\-component2\_fraction}) of the total $^{56}$Ni mass is placed uniformly over $q_{\mathrm{mix}} < q \leq q_{\mathrm{mix,2}}$, and the remaining $(1 - f_{\mathrm{outer}})$ is placed uniformly over $0 \leq q \leq q_{\mathrm{mix}}$, where $q_{\mathrm{mix,2}}$ (\texttt{Ni\_mix\_\-component2\_extent}) sets the outer boundary of the second component. This produces a step-function Ni profile with two distinct concentration levels: a denser core and a dilute outer shell. This kernel is motivated by multi-dimensional explosion simulations that show fingers of $^{56}$Ni reaching significantly farther into the ejecta envelope than the bulk of the synthesized material \citep{Kifonidis2003}. If the outer-shell support mass is zero (e.g., because $q_{\mathrm{mix,2}} \leq q_{\mathrm{mix}}$), the outer Ni mass is folded back into the core component. Setting $f_{\mathrm{outer}} = 0$ recovers Kernel~1 behavior.
\end{itemize}

A key implementation difference is that all Ni placement and boxcar smoothing are performed on the high-resolution progenitor profile \textit{before} mapping to the simulation grid. This preserves composition structure that would be lost if mixing were done on the coarse simulation grid for low zone-count simulations. The processed profile is stored in memory and reused whenever the grid is remapped (Sect.~\ref{sec:remap}).

During composition normalization, the Ni mass fraction is decoupled from the other species: if the total mass fractions sum to more than 1 after Ni placement, the non-Ni species are rescaled downward to accommodate the Ni without changing $X_{\mathrm{Ni}}$ itself. This prevents the normalization step from diluting the intended Ni mass. When \texttt{Ni\_by\_hand\,=\,1}, the progenitor's native $^{56}$Ni abundance is zeroed before seeding the kernel profile, ensuring that the on-grid Ni mass is entirely controlled by \texttt{Ni\_mass} and the chosen kernel shape.

\subsubsection{Independent bulk ejecta and $^{56}$Ni profile smoothing}
\label{sec:ni_independent_smoothing}

Progenitor composition profiles from stellar evolution codes often contain sharp zone-to-zone discontinuities that benefit from smoothing before use in a radiation-hydrodynamics simulation. In SNEC, this is done via boxcar smoothing with a single hard-coded mass window and iteration count applied to all composition species simultaneously, coupling the $^{56}$Ni smoothing scale to that of the bulk composition. SuperSNEC exposes both the bulk smoothing parameters (\texttt{boxcar\_\-nominal\_\-mass\_\-msun}, \texttt{boxcar\_\-number\_\-iterations}) and independent $^{56}$Ni smoothing parameters (\texttt{boxcar\_ni\_\-nominal\_mass\_msun}, \texttt{boxcar\_ni\_\-number\_iterations}), allowing the user to smooth the bulk composition and the Ni profile on different scales. The independent smoothing of $^{56}$Ni coupled with the three available mixing kernels (Sect.~\ref{sec:ni_mixing}) enables a high level of flexibility in producing the final $^{56}$Ni profile of the ejecta.

\subsection{Photosphere Ni luminosity smoothing}
\label{sec:ni_smoothing}

In both SNEC and SuperSNEC, the observed bolometric luminosity $L_\mathrm{obs}$ is computed as the sum of the photospheric luminosity $L_\mathrm{photo}$ and the $^{56}$Ni heating deposited in all zones above the photosphere:
\begin{equation}
L_\mathrm{obs} = L_\mathrm{photo} + \sum_{i \ge i_\mathrm{photo}} \dot{q}_\mathrm{Ni}(i)\, \Delta m_i,
\end{equation}
where $i_\mathrm{photo}$ is the zone index of the photosphere ($\tau = 2/3$), $\dot{q}_\mathrm{Ni}(i)$ is the specific Ni heating rate in zone~$i$, and $\Delta m_i$ is the zone mass. Because the sum starts at a discrete zone boundary rather than the exact (interpolated) photosphere location, the Ni contribution exhibits small step-like jumps each time $i_\mathrm{photo}$ decreases by one as the photosphere recedes inward.

At 1000 zones these steps are negligible, but at ${\sim}100$ zones each zone contains a larger fraction of the Ni mass and the steps can produce ${\sim}1$--$2\%$ luminosity discontinuities. We remove this artifact by adding a fractional contribution from the zone immediately below the photosphere. The photosphere mass coordinate $m_\mathrm{photo}$ (interpolated at $\tau = 2/3$) lies between zones $i_\mathrm{photo}-1$ and $i_\mathrm{photo}$. The fraction of zone $i_\mathrm{photo}-1$ above the photosphere,
\begin{equation}
f = \frac{m(i_\mathrm{photo}) - m_\mathrm{photo}}{m(i_\mathrm{photo}) - m(i_\mathrm{photo}-1)},
\end{equation}
ramps smoothly from 0 to 1 as the photosphere traverses the zone. The corrected luminosity is $L_\mathrm{corr} = L_\mathrm{obs} + f \times \dot{q}_\mathrm{Ni}(i_\mathrm{photo}-1)\,\Delta m_{i_\mathrm{photo}-1}$. This correction is only applied when the potential step exceeds 2\% of $L_\mathrm{obs}$; at early times the zones are thin and the hard cutoff is already accurate. The smoothing is controlled by \texttt{smooth\_ni\_\-luminosity} (1\,=\,on, 0\,=\,off; default on).

\begin{figure*}
\centering
\includegraphics[width=0.98\linewidth]{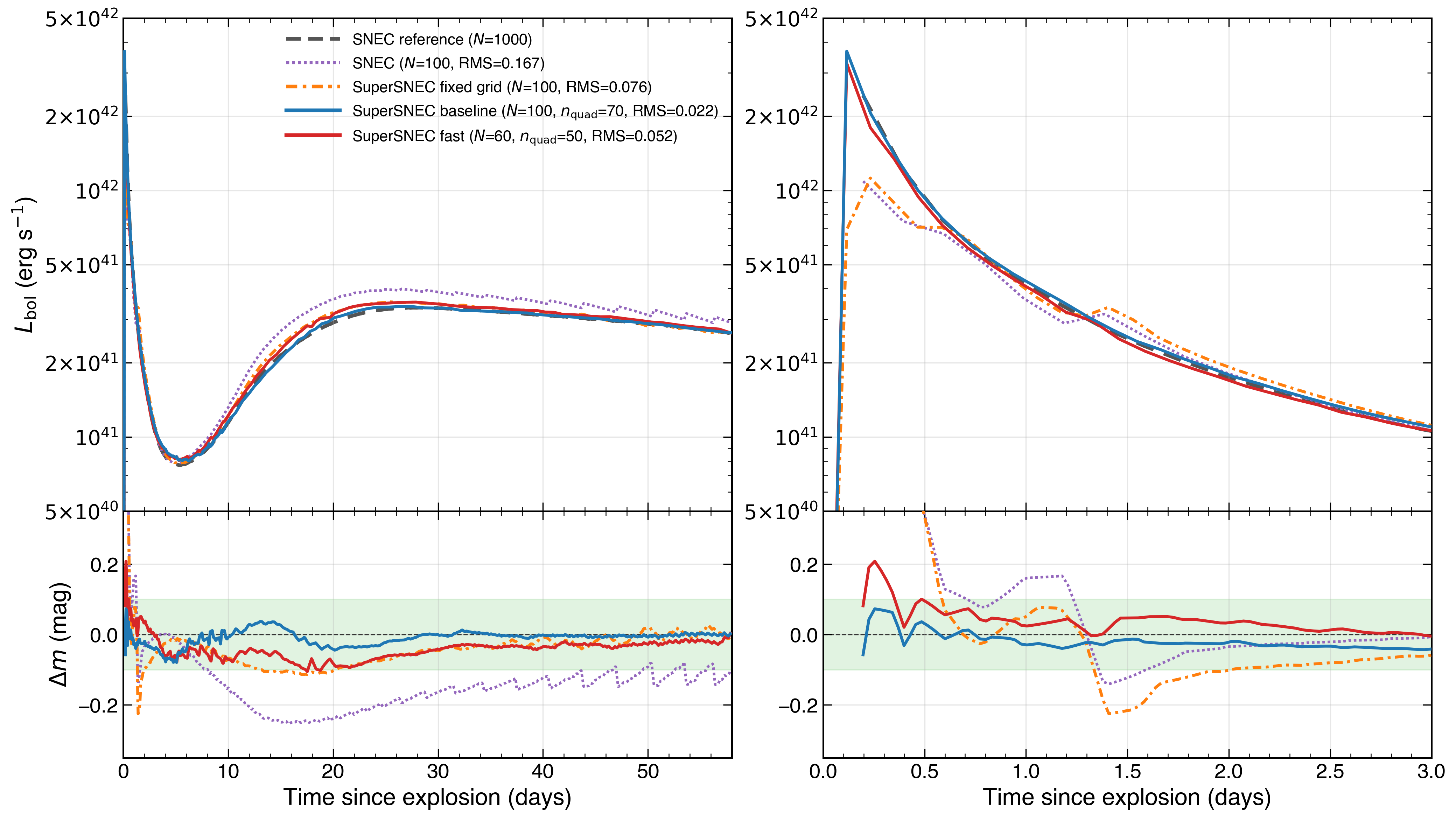}
\caption{Bolometric light-curve comparison of five configurations against a 1000-zone SNEC reference (black dashed line). \textit{Top panels:} luminosity; \textit{bottom panels:} magnitude residuals $\Delta m = -2.5\log_{10}(L/L_\mathrm{ref})$, with the green band marking $\pm0.1$~mag. \textit{Left:} full evolution (0--58~d); \textit{right:} early-time zoom (0--3~d). The original SNEC code at $N=100$ (purple dotted; $\Delta m_\mathrm{all}=0.167$~mag) diverges strongly at $t\lesssim0.5$~d and $t\gtrsim10$~d. SuperSNEC with the same fixed computational grid (orange dot-dashed) improves via adaptive \Nif\ cadence and solver tuning. The adaptive-runtime grid brings a further factor-of-three improvement: the baseline ($N=100$, $n_\mathrm{quad}=70$; blue solid line) and fast ($N=60$, $n_\mathrm{quad}=50$; red solid line) configurations both remain well within 0.1~mag across all phases. The SuperSNEC fixed-grid model uses $f_\mathrm{Ni}=0.10$; the baseline uses $f_\mathrm{Ni}=0.20$, $n_\mathrm{quad}=70$; the fast model uses $f_\mathrm{Ni}=0.20$, $n_\mathrm{quad}=50$. All three share $\alpha_\mathrm{surf}=7$, $t_\mathrm{relax}=5$~d, $\Delta t_{\mathrm{Ni,min}}=5\times10^4$~s, $\Delta t_{\mathrm{Ni,max}}=5.456\times10^5$~s. The original SNEC uses $\Delta t_{\mathrm{Ni,min}}=5\times10^4$~s with no adaptive cadence.}
\label{fig:adaptive_runtime}
\end{figure*}

\section{Results}
\label{sec:validation}
To validate our runtime oriented optimizations and the adaptive runtime mode, we use comparisons to bolometric light curves from a 1000-zone reference SNEC run with default SNEC settings ($\mathrm{EPSTOL}=10^{-7}$, $\mathrm{ITMAX}(100,300)$, fixed computational grid from \texttt{Grid\-Pattern.dat}). The presupernova model is the ``He4'' stripped-envelope helium star from \citet{Morozova2015}, generated with the stellar evolution code MESA \citep{Paxton2011} from a $15\,\Msun$ zero-age main-sequence progenitor whose hydrogen envelope was removed during binary interaction \citep{PiroMorozova2014}. This model ships with the SNEC distribution; any presupernova profile in the same format (radial structure plus isotopic composition) is compatible with SuperSNEC. We use the thermal bomb energy injection mode throughout this work, since the final kinetic energy is a direct input parameter, unlike the piston mode where it must be inferred from piston velocity and timing. The reference runtime is $671$~seconds on an Apple Silicon M1~Pro CPU.

A 1000-zone SuperSNEC run using the \texttt{legacy\_\-pattern} grid mode, our new $^{56}$Ni mixing prescription (Sect.~\ref{sec:ni_mixing}), the relaxed solver tolerance ($\mathrm{EPSTOL} = 10^{-4}$; Sect.~\ref{sec:solver}), and a conservative $^{56}$Ni update cadence ($f_{\mathrm{Ni}} = 0.10$, comparable to the original SNEC fixed-interval updates) reproduces this reference to RMS~$= 0.017$~mag across all phases. This confirms that, when the grid and deposition cadence are kept close to the original SNEC configuration, the remaining code changes (mixing prescription, solver tolerance) do not measurably bias the light curve. Reported metrics are the RMS magnitude residual $\Delta m = -2.5\log_{10}(L_{\mathrm{run}}/L_{\mathrm{ref}})$ evaluated in three phase bins: 0--5~d, 5--30~d, and $>30$~d. 

\subsection{Adaptive Runtime Validation}
\label{sec:adaptive_validation}

Figure~\ref{fig:adaptive_runtime} compares bolometric light curves from five configurations against a 1000-zone SNEC reference LC. The original, unmodified SNEC code at $N=100$ zones (purple dotted line) produces a qualitatively similar LC, but with some significant deviations. The overall LC deviation when compared to the 1000-zone reference is $\Delta m_\mathrm{all}=0.167$~mag, which is driven by a ${\sim}0.3$~mag systematic around $t\approx 20\text{--}40$~d where the photosphere recedes into the poorly resolved interior. Running the same 100-zone fixed grid through SuperSNEC, which adds the relaxed solver tolerance ($\mathrm{EPSTOL}=10^{-4}$; Sect.~\ref{sec:solver}), adaptive $^{56}$Ni deposition cadence (Sect.~\ref{sec:ni_cadence}), and increased maximum timestep ($\Delta t_{\mathrm{max}}=10^5$~s), reduces the residual to $\Delta m_\mathrm{all}=0.076$~mag (orange dot-dashed line), without enabling the adaptive grid. Enabling the adaptive-runtime grid (Sect.~\ref{sec:gridding}) brings a further improvement: the SuperSNEC baseline ($N=100$, $n_\mathrm{quad}=70$, Sect.~\ref{sec:baseline}) achieves $\Delta m_\mathrm{all}=0.022$~mag in ${\sim}1.6$~s. The fastest configuration ($N=60$, $n_\mathrm{quad}=50$) shows $0.052$~mag in ${\sim}0.8$~s, still better than the 100-zone fixed grid.

The adaptive grid mode (Sect.~\ref{sec:gridding}) has two user-facing parameters: the surface concentration $\alpha_{\mathrm{surf}}$ and the relaxation timescale $t_{\mathrm{relax}}$. The defaults ($\alpha_{\mathrm{surf}} = 7$, $t_{\mathrm{relax}} = 5$~d) were validated through the zone-count scaling scan in Table~\ref{tab:imax_scaling}, which shows monotonically improving RMS from $N = 60$ to $1000$, and through the five-model comparison in Figure~\ref{fig:adaptive_runtime}. The inner fraction $f_{\mathrm{inner}}$ and the photosphere-scaled $\alpha$ are derived quantities (Sect.~\ref{sec:gridding}) and do not require tuning.

Both the baseline and fast adaptive SuperSNEC configurations are suitable for large model-grid applications. The largest deviations for both models occur at very early times ($t \lesssim 0.3$~d), where the baseline model peaks at ${\sim}0.07$~mag and the fast model briefly reaches ${\sim}0.2$~mag; both settle well within $0.1$~mag for all subsequent phases.

\subsection{Cumulative Speedup}
\label{sec:speed_strategy}
\label{sec:cumulative_speedup}
\label{sec:speed_results}
Both SNEC and SuperSNEC runtimes strongly depend on configuration parameter settings. In SuperSNEC we choose to expose more configurable parameters to the user to allow for problem-specific runtime and accuracy optimization. We determined optimal values for the dominant runtime levers across both the default SNEC parameters and the new parameters added by SuperSNEC (Table~\ref{tab:parameters}) through parameter scans, locking each optimized setting before scanning the next:
\begin{itemize}
\item Solver: $\mathrm{EPSTOL}$ and $\mathrm{ITMAX}$ (Appendix~\ref{sec:solver_sweep}).
\item Quadrature: $n_\mathrm{quad}$ (Appendix~\ref{sec:ni_quad}).
\item $^{56}$Ni cadence: $f_{\mathrm{Ni}}$, $\Delta t_{\mathrm{Ni,min}}$, $\Delta t_{\mathrm{Ni,max}}$ (Appendix~\ref{sec:ni_cadence_sweep}).
\item Remap cadence: $\Delta t_{\mathrm{remap}}$ (Appendix~\ref{sec:grid_cadence_sweep}).
\item Maximum timestep: $\Delta t_{\mathrm{max}}$ (Appendix~\ref{sec:dtmax_sweep}).
\end{itemize}
Output mode (\texttt{output\_mode}; Appendix~\ref{sec:output_cadence_speed}) was benchmarked separately as a production-grid option. Parameters not yet scanned in isolation but with potential impact include \texttt{opacity\_floor\_\-envelope} and \texttt{opacity\_floor\_core} (affect diffusion timescale and timestep selection; also nuisance parameters for light-curve inference). The Saha ionization species count \texttt{saha\_ncomps} was tested separately: reducing below the default of 3 degrades light-curve quality significantly, while increasing beyond 3 adds runtime with no measurable accuracy gain, so the default is already optimal.

Combined with zone reduction ($1000 \to 100$ zones) and the code updates described in Sect.~\ref{sec:code_updates}, these parameter optimizations compound into a total speedup of ${\sim}420\times$ relative to a 1000-zone SNEC reference run. Table~\ref{tab:cumulative_speedup} demonstrates this by applying each optimization successively: each row enables one additional change on top of all previous rows. The ``Cumulative'' column gives the total speedup factor relative to the SNEC reference; the ``Marginal'' column shows the incremental gain from that step alone. 

\begin{deluxetable*}{lrccc}
\tabletypesize{\small}
\tablecaption{Cumulative speedup breakdown. Each row adds one optimization on top of the previous row. Runtime is wall-clock time on an M1~Pro Apple Silicon CPU. RMS is computed against the 1000-zone reference.\label{tab:cumulative_speedup}}
\tablehead{
\colhead{Configuration} &
\colhead{Runtime (s)} &
\colhead{Cumulative} &
\colhead{Marginal} &
\colhead{$\Delta m_\mathrm{all}$ (mag)}
}
\startdata
SNEC reference (1000 zones) & 671.0 & $1\times$ & --- & --- \\
\hline
Zone reduction ($N{=}100$) & 10.4 & $64\times$ & $64.3\times$ & 0.167 \\
Compiler flags (\texttt{-Ofast}) & 8.6 & $78\times$ & $1.2\times$ & 0.167 \\
SuperSNEC codebase & 8.4 & $80\times$ & $1.0\times$ & 0.074 \\
Solver tolerance ($10^{-7} \to 10^{-4}$) & 7.2 & $94\times$ & $1.2\times$ & 0.075 \\
Ni ray-tracing optimization & 3.3 & $201\times$ & $2.1\times$ & 0.076 \\
Ni cadence ($f_{\mathrm{Ni}}{=}0.10$) & 3.0 & $222\times$ & $1.1\times$ & 0.075 \\
Quadrature ($150 \to 70$) & 1.7 & $396\times$ & $1.8\times$ & 0.077 \\
Adaptive grid + $f_{\mathrm{Ni}}{=}0.20$ & 1.6 & $422\times$ & $1.1\times$ & 0.022 \\
\hline
\textbf{Total (SuperSNEC baseline)} & \textbf{1.6} & $\mathbf{422\times}$ & --- & \textbf{0.022} \\
\hline
Fast ($N{=}60$, $n_{quad}{=}50$) & 0.8 & $845\times$ & $2.0\times$ & 0.052 \\
\hline
\\
\multicolumn{5}{c}{\textit{Production grid mode (\texttt{output\_mode=1}; Sect.~\ref{sec:output_mode})}} \\
\hline
Baseline & 1.3 & $509\times$ & $1.2\times$ & 0.022 \\
Fast & 0.6 & $1090\times$ & $1.3\times$ & 0.052 \\
\enddata
\tablecomments{The reference is unmodified SNEC~1.01 with 1000 zones, compiled with \texttt{-O3 -fbounds-check}, \texttt{EPSTOL}$=10^{-7}$, and fixed \Nif\ cadence ($\Delta t_\mathrm{Ni}=5\times10^4$~s). All SuperSNEC runs use 100 zones unless otherwise noted. All rows above the production grid section use \texttt{output\_mode=0} (full SNEC output). Production grid rows apply \texttt{output\_mode=1} to their respective configurations; the marginal column shows the speedup from enabling minimal output relative to the full-output row.}
\end{deluxetable*}

The dominant factor is zone reduction, which alone provides a ${\sim}64\times$ speedup. Compiler optimization (\texttt{-Ofast}) adds another ${\sim}1.2\times$. Switching to the SuperSNEC codebase, which includes the diagnostic cadence optimization (Sect.~\ref{sec:cadence}) and composition handling improvements (Sect.~\ref{sec:ni_mixing}), does not measurably change runtime at these settings (the solver dominates at $\mathrm{EPSTOL}=10^{-7}$ with full quadrature) but improves accuracy from 0.167 to 0.074~mag RMS. Relaxing the solver tolerance from $10^{-7}$ to $10^{-4}$ (Sect.~\ref{sec:solver}) provides ${\sim}1.2\times$, and the Ni ray-tracing optimization (Sect.~\ref{sec:ni_ray_interp}) yields a further ${\sim}2.1\times$. The adaptive Ni cadence and quadrature reduction collectively add another ${\sim}2\times$. Finally, enabling the adaptive grid mode (Sect.~\ref{sec:gridding}) with $f_{\mathrm{Ni}} = 0.20$ provides a modest additional speedup (${\sim}1.1\times$) and is the key enabler that makes 100 zones \emph{accurate}, reducing RMS from ${\sim}0.08$~mag (fixed grid) to ${\sim}0.022$~mag. If a somewhat higher RMS (${\sim}0.05$~mag) can be tolerated, another ${\sim}2\times$ speedup can be achieved with 60 zones and $n_\mathrm{quad} = 50$ (``fast'' mode).

For large model-grid production runs aimed at LC fitting, the minimal output mode (Sects.~\ref{sec:output_mode} and~\ref{sec:output_cadence_speed}) provides an additional ${\sim}1.2\times$ speedup by writing only the 5 output files needed by LC fitting codes instead of the standard 61 SNEC output files. This reduces the baseline runtime from $1.6$~s to ${\sim}1.3$~s per model (${\sim}510\times$ faster than the SNEC reference run), or ${\sim}0.6$~s (${\sim}1090\times$) with the fast configuration. A million-model parameter-space sweep can thus be completed in ${\sim}15$~d of single-core wall-clock time with the baseline, or ${\sim}7$~d with the fast configuration. Because each model is independent and its runtime does not saturate a single core, trivial parallelism (launching independent SNEC instances) scales near-linearly with available cores: on a 16-core workstation, the same million-model grid should complete in ${\sim}1$~d (baseline) or ${\sim}10$~h (fast).

\subsection{Working Baseline}
\label{sec:baseline}

Based on the validation of the adaptive gridding mode (Sect.~\ref{sec:adaptive_validation}) and the parameter scans described above (detailed in Appendix~\ref{sec:speed_details}), we define recommended baseline settings for 100-zone \texttt{adaptive\_\-runtime} models:
\begin{itemize}
\item Solver: $\mathrm{EPSTOL} = 10^{-4}$, $\mathrm{ITMAX} = (100, 300)$.
\item $^{56}$Ni cadence: $\Delta t_{\mathrm{Ni,min}} = 5.0\times10^4$~s, $\Delta t_{\mathrm{Ni,max}} = 5.456\times10^5$~s, $f_{\mathrm{Ni}} = 0.20$.
\item Remap cadence: $\Delta t_{\mathrm{remap}} = 1.0$~d, $q_\mathrm{stop} = 0.50$ (Sect.~\ref{sec:remap_scheduling}).
\item Grid: $\alpha_{\mathrm{surf}} = 7.0$, $t_{\mathrm{relax}} = 5.0$~d (Sect.~\ref{sec:gridding}).
\item Quadrature: $n_\mathrm{quad} = 70$ (Appendix~\ref{sec:ni_quad}).
\item Timestep ceiling: $\Delta t_{\mathrm{max}} = 10^5$~s ($10\times$ the SNEC default).
\item Ni luminosity smoothing: on (Sect.~\ref{sec:ni_smoothing}).
\item Output mode: full (Sect.~\ref{sec:output_mode}). For production grid runs, minimal output is recommended.
\end{itemize}
The complete list of all configuration parameters and their baseline values is given in Table~\ref{tab:parameters}. With these settings the baseline achieves $\Delta m_\mathrm{all} = 0.022$~mag at $N = 100$ (Figure~\ref{fig:adaptive_runtime}); the corresponding runtimes and cumulative speedup factors are summarized in Table~\ref{tab:cumulative_speedup}.

\begin{figure*}[t!]
\centering
\includegraphics[width=\linewidth]{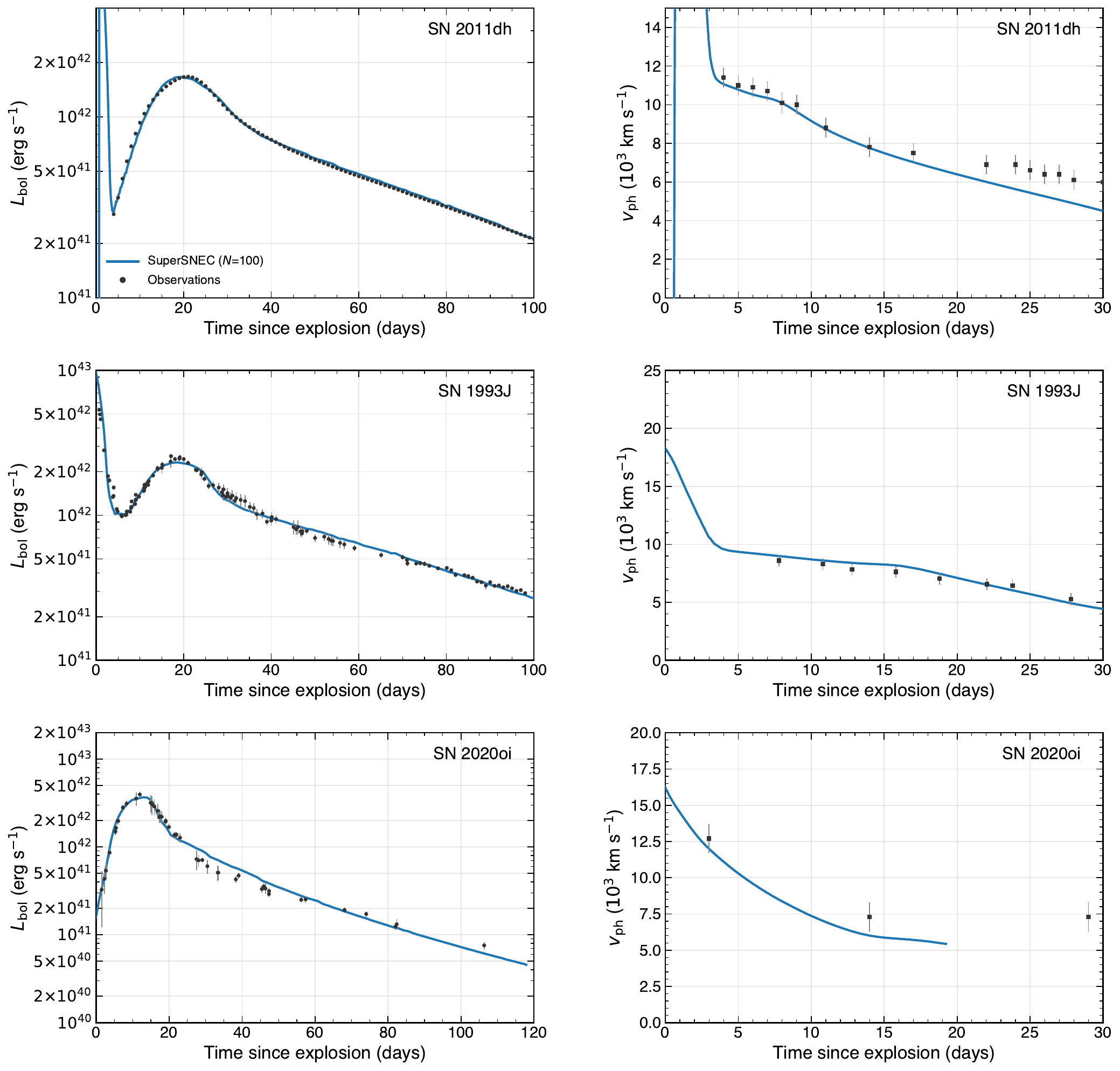}
\caption{Bolometric light-curve (left) and photospheric velocity (right) comparison of $N$=100 SuperSNEC models (blue lines) to SN~2011dh (top), SN~1993J (middle), and SN~2020oi (bottom) data (black markers). Velocities for SN~2011dh and SN~1993J are Fe\,\textsc{ii}\,$\lambda5169$ absorption-minimum measurements; SN~2020oi uses Si\,\textsc{i}\,$\lambda10460$ as Fe\,\textsc{ii}\,$\lambda5169$ could not be reliably measured. Observational data: SN~2011dh bolometric LC from \citet{Ergon2014,Ergon2015}, velocities from \citet{Marion2014}; SN~1993J bolometric LC constructed from \citet{Richmond1994} photometry with UV/IR blackbody corrections, velocities measured from UNLV Supernova Archive spectra \citep{Barbon1995}; SN~2020oi bolometric LC constructed from the \citet{Rho2021} optical photometry using a blackbody bolometric correction, with Si\,\textsc{i}\,$\lambda10460$ velocities also from \citet{Rho2021}.}
\label{fig:lc_comp}
\end{figure*}

\subsection{Test cases: SN~2011dh, SN~1993J, and SN~2020oi}
\label{sec:lc_comp}

As a demonstration that low-zone-count SuperSNEC models can reproduce observed stripped-envelope supernova light curves, we compare 100-zone models to three well-studied events: two canonical SNe~IIb, SN~2011dh \citep{Ergon2014} and SN~1993J \citep{Richmond1994}, and one SN~Ic, SN~2020oi \citep{Rho2021}. Table~\ref{tab:lc_comp} summarizes the progenitor and explosion parameters used in our models alongside literature values.

These models were not found via thorough parameter space exploration or fitting; they are simply hand-picked models based on pre-supernova progenitor models drawn from a grid of binary MESA \citep{Paxton2011,Paxton2013} stellar-evolution models for stripped-envelope SNe (Hinds et al., in preparation). Each binary system (characterized by primary mass $M_1$, secondary mass $M_2$, and initial orbital period $P$) is evolved through Roche-lobe overflow and envelope stripping to the onset of core collapse, producing a He-star with a residual hydrogen envelope. For Type~IIb models, the resulting pre-supernova star is described by its stripped-core mass $M_\mathrm{core}$ (the total pre-supernova mass excluding any residual hydrogen envelope of mass $M_\mathrm{H}$), and pre-supernova radius $R$. For the Type~Ic model (SN~2020oi), the hydrogen envelope was fully removed to produce a compact, hydrogen-free progenitor. The explosion is parametrized by the final kinetic energy $E_k$, synthesized $^{56}$Ni mass $M_{^{56}\mathrm{Ni}}$, and the ejecta mass $M_\mathrm{ej}$ (total pre-supernova mass minus the compact-remnant mass).

\paragraph{SN~2011dh}
The top panels of Figure~\ref{fig:lc_comp} compare the SuperSNEC model to the observed bolometric LC and photospheric expansion velocity evolution of SN~2011dh~\citep{Ergon2014}. The progenitor is a $M_1 = 16~\Msun$, $M_2 = 14~\Msun$ binary ($P = 160$~d) that produces a pre-supernova star with $M_\mathrm{core} = 4.28~\Msun$, $R = 100~\Rsun$, and $M_\mathrm{H} = 0.046~\Msun$. We explode this model with $E_k = 1.0 \times 10^{51}$~erg and $M_{^{56}\mathrm{Ni}} = 0.074~\Msun$, yielding an ejecta mass of $M_\mathrm{ej} = 2.45~\Msun$ (compact-remnant mass $1.98~\Msun$). The $^{56}$Ni mixing uses a step-function kernel with $q_\mathrm{mix} = 0.45$. The opacity floor is $0.03$~cm$^2$~g$^{-1}$ in the envelope and $0.12$~cm$^2$~g$^{-1}$ in the core\footnote{These opacity floors are treated as tunable numerical parameters and adjusted by hand to improve agreement between the model and the observed light curve; they are not intended as direct physical measurements of the true ejecta opacity.}
. As shown in Table~\ref{tab:lc_comp}, our explosion parameters are broadly consistent with literature values.

\paragraph{SN~1993J}
The middle panels of Figure~\ref{fig:lc_comp} show an analogous comparison for SN~1993J, using a $M_1 = 15~\Msun$, $M_2 = 13~\Msun$ binary ($P = 240$~d) progenitor with $M_\mathrm{core} = 3.95~\Msun$, $R = 428~\Rsun$, and $M_\mathrm{H} = 0.10~\Msun$. We explode this model with $E_k = 1.35 \times 10^{51}$~erg and $M_{^{56}\mathrm{Ni}} = 0.10~\Msun$, yielding $M_\mathrm{ej} = 2.56~\Msun$ (compact-remnant mass $1.63~\Msun$). The $^{56}$Ni mixing uses $q_\mathrm{mix} = 0.17$ with a secondary component fraction of $0.25$ extending to $q = 1.0$. The opacity floor is $0.03$~cm$^2$~g$^{-1}$ in the envelope and $0.06$~cm$^2$~g$^{-1}$ in the core. These parameters fall within the literature range for SN~1993J (Table~\ref{tab:lc_comp}). The photospheric velocities are Fe\,\textsc{ii}\,$\lambda5169$ absorption-minimum measurements from spectra in the UNLV Supernova Spectrum Archive \citep{Barbon1995}, extracted by fitting the trough minimum in the 5000--5120~\AA\ window. The bolometric light curve we compare against is constructed from the \citet{Richmond1994} photometry with UV and IR blackbody corrections at $E(B-V) = 0.08$~mag.

\paragraph{SN~2020oi}
The bottom panels of Figure~\ref{fig:lc_comp} show the comparison for SN~2020oi, a Type~Ic supernova in M100 \citep{Rho2021}. We use a compact, hydrogen-free progenitor derived from a $M_1 = 11~\Msun$, $M_2 = 9~\Msun$ binary ($P = 280$~d) model from which the residual hydrogen envelope has been removed. Unlike the pure CO progenitor adopted by \citet{Rho2021}, our model retains a helium envelope. This is not obviously inconsistent with SN~2020oi, for which \citet{Rho2021} discuss possible He\,\textsc{i} signatures in the optical and near-IR spectra; similar helium-hidden Ic interpretations have also been explored by \citet{Dessart2012}. The truncated hydrodynamic model has $M_\mathrm{core} = 1.97~\Msun$ and $R = 1.1~\Rsun$. With $1.00~\Msun$ excised as a compact remnant, the ejecta mass is $M_\mathrm{ej} = 0.97~\Msun$. We explode this model with $E_k = 0.80 \times 10^{51}$~erg and $M_{^{56}\mathrm{Ni}} = 0.080~\Msun$. The $^{56}$Ni mixing uses a two-component kernel ($q_\mathrm{mix} = 0.07$) with a secondary component fraction of $0.19$ extending to $q = 0.8$. The opacity floor is $0.01$~cm$^2$~g$^{-1}$ in the envelope and $0.06$~cm$^2$~g$^{-1}$ in the core. The bolometric light curve is constructed from the \citet{Rho2021} optical photometry with blackbody bolometric corrections at $E(B-V) = 0.199$~mag, matching the reddening adopted by \citet{Rodriguez2024}. The photospheric velocity tracer for SN~2020oi is Si\,\textsc{i}\,$\lambda10460$, as Fe\,\textsc{ii}\,$\lambda5169$ could not be reliably measured in the spectra. The photospheric velocity evolution in the model for SN~2020oi terminates earlier than for the two SNe~IIb because the low ejecta mass causes the photosphere to recede through the ejecta more rapidly, rendering it undefined at earlier times.

All three SuperSNEC models used 100-zone computational grids with the adaptive runtime gridding settings described in Section~\ref{sec:gridding} and the working baseline parameters from Section~\ref{sec:baseline} (Table~\ref{tab:parameters}). The models reproduce the bolometric light curves and photospheric velocity evolution of each supernova, for explosion parameters broadly consistent with the literature values (Table~\ref{tab:lc_comp}).

\begin{deluxetable*}{lcccccc}
\setlength{\tabcolsep}{4pt}
\tablecaption{SuperSNEC model parameters compared to literature values for SN~2011dh, SN~1993J, and SN~2020oi.\label{tab:lc_comp}}
\tablewidth{0pt}
\tablehead{
\colhead{Parameter} &
\multicolumn{2}{c}{SN~2011dh (IIb)} &
\multicolumn{2}{c}{SN~1993J (IIb)} &
\multicolumn{2}{c}{SN~2020oi (Ic)} \\
\colhead{} &
\colhead{This work} &
\colhead{Literature} &
\colhead{This work} &
\colhead{Literature} &
\colhead{This work} &
\colhead{Literature}
}
\startdata
$R$ ($\Rsun$)                            & \textbf{100}   & 200--280\tablenotemark{a}    & \textbf{428}   & 430--620\tablenotemark{b} & \textbf{1.1}    & \nodata \\
$M_\mathrm{core}$ ($\Msun$)               & \textbf{4.28}  & 3.1--4.0\tablenotemark{a}    & \textbf{3.95}  & 3.0--5.0\tablenotemark{b} & \textbf{1.97}    & 2.16\tablenotemark{d} \\
$M_\mathrm{H}$ ($\Msun$)                & \textbf{0.046} & 0.01--0.10\tablenotemark{a}  & \textbf{0.10}  & 0.1--0.4\tablenotemark{b} & \textbf{0}       & 0 \\
$M_\mathrm{ej}$ ($\Msun$)               & \textbf{2.45}  & 1.7--2.5\tablenotemark{a}    & \textbf{2.56}  & 1.9--3.5\tablenotemark{b} & \textbf{0.97}    & 0.7--1.1\tablenotemark{c} \\
$E_k$ ($10^{51}$~erg)                   & \textbf{1.0}   & 0.6--1.0\tablenotemark{a}    & \textbf{1.35}  & 1.0--1.6\tablenotemark{b} & \textbf{0.80}    & 0.7--0.9\tablenotemark{c} \\
$M_{^{56}\mathrm{Ni}}$ ($\Msun$)        & \textbf{0.074} & 0.06--0.09\tablenotemark{a}  & \textbf{0.10}  & 0.06--0.14\tablenotemark{b} & \textbf{0.080} & 0.05--0.09\tablenotemark{c} \\
\enddata
\tablenotetext{a}{\citet{Bersten2012}; \citet{Ergon2014}; \citet{Ergon2015}; \citet{Marion2014}; \citet{Jerkstrand2015}.}
\tablenotetext{b}{\citet{Woosley1994}; \citet{Shigeyama1994}; \citet{Young1995}; \citet{Maund2004}.}
\tablenotetext{c}{\citet{Rho2021}; \citet{Gagliano2022}.}
\tablenotetext{d}{Pre-SN mass from \citet{Rho2021}; their CO progenitor model is He free.}
\end{deluxetable*}

\section{Discussion}
\label{sec:discussion}

For all three SNe we have modeled here the 100-zone SuperSNEC models reproduce the observed bolometric light curves and photospheric velocity evolution well (Figure~\ref{fig:lc_comp}). It could even be argued that the agreement between the data and the models is remarkable, given the simplified 1D radiation-hydrodynamics treatment, the very low zone count ($N = 100$), and the relatively simple parametric $^{56}$Ni mixing prescription. We obtained these matches using hand-picked models with manually tuned explosion parameters, without scanning a large model grid. We do not attempt to derive formal uncertainty intervals for the explosion parameters here, as these comparisons are intended to validate the use of SuperSNEC for constructing large parameter-space model grids. 

None of the three objects shows any significant systematic luminosity excess above the models at any phase. SN~2020oi is of particular interest in the context of the analysis by \citet{Rodriguez2024}, who applied the Katz-integral method to a sample of stripped-envelope SNe and argued that many events require a power source in addition to $^{56}$Ni decay. In their Extended Data Tables~1--3, SN~2020oi is the strongest apparent-excess case: they report $L_{T-nuc}/L_{T100} = 0.388 \pm 0.026$ and infer $M_{^{56}\mathrm{Ni}} \approx 0.027~\Msun$, adopting a total reddening of $E(B-V) = 0.199$~mag (Galactic plus host). We adopt the same reddening for our bolometric light-curve construction (Section~\ref{sec:lc_comp}). However, the Katz-integral method infers a $^{56}$Ni mass roughly a factor of three lower than the $0.080~\Msun$ in our model. Our purely radioactive SuperSNEC model reproduces the observed bolometric light curve of SN~2020oi from ${\sim}1.5$ to ${\sim}106$~d without any non-radioactive power source. Rather than signaling a missing power source, the large Katz-integral ``excess'' reflects the sensitivity of the inferred $^{56}$Ni mass to assumptions about progenitor structure and mixing geometry. We note that radio observations of SN~2020oi do indicate the presence of circumstellar material \citep{Horesh2020,Maeda2021}, but the CSM interaction does not appear to contribute significantly to the optical luminosity over the first ${\sim}100$~d covered by the bolometric LC built from optical data.

We emphasize that some stripped-envelope SNe do show unambiguous signatures of additional power sources beyond $^{56}$Ni decay. Late-time re-brightening from CSM interaction has been observed in several SE SNe, for example SN~2019oys and SN~2019tsf \citep{Sollerman2020}, and SN~2014C underwent a spectral metamorphosis from Type~Ib to Type~IIn as its ejecta collided with a hydrogen-rich shell \citep{Milisavljevic2015}. These events are morphologically distinct, with prominent light-curve bumps, re-brightening episodes, or spectral changes that clearly signal non-radioactive power. Even in less extreme cases, lower levels of CSM interaction that do not produce obvious re-brightening should still manifest as a systematic luminosity excess above a purely radioactive model at some epoch. SN~2020oi shows none of these features: its bolometric light curve is smooth and monotonically declining after peak, with no systematic excess above our model at any phase where we have data coverage, consistent with pure $^{56}$Ni-powered diffusion.

In a broader context, \citet{Afsariardchi2021} examined 27 stripped-envelope SNe and found that applying Arnett's rule \citep{Arnett1982} at peak yields $M_{^{56}\mathrm{Ni}}$ estimates typically ${\sim}2$ times larger than tail-based values, a tension they suggest could be alleviated if ${\sim}7$--$50\%$ of the peak luminosity were supplied by an additional power source (e.g., CSM interaction or magnetar spin-down). Our three test cases show that such discrepancies can also arise from the sensitivity of analytic $^{56}$Ni estimates to progenitor structure and mixing geometry, without requiring any power source beyond radioactive decay and the hydrodynamics of the explosion itself.

Three modeled events do not settle a population-level question, but the low per-model runtime of SuperSNEC makes it practical to test this explanation systematically via large model-grid inference. Events for which a systematic luminosity excess persists even after matching to radiation-hydrodynamic models would constitute stronger evidence for truly additional power sources; CSM interaction could be tested by including circumstellar material in the presupernova profiles, while a central engine could be added as an alternative heating module.

\subsection{Further optimization opportunities}

A few additional optimization opportunities exist in the SNEC codebase that we have not yet pursued.

\textit{Saha ionization caching.} The Paczynski equation of state calls the simplified Saha solver (\texttt{simple\_saha.F90}) at every zone, every Newton--Raphson iteration, every timestep. At $N=100$ zones with $\mathrm{EPSTOL}=10^{-4}$, the two implicit solvers (hydro and radiative transfer) each require ${\sim}3$--5 iterations per timestep, yielding ${\sim}600$--$1000$ Saha evaluations per timestep and ${\sim}10^6$ over a full simulation. Each evaluation solves a Newton--Raphson ionization balance for 3 species (\texttt{saha\_ncomps\,=\,3}); the remaining 16 composition species take the fully-ionized fast path. Reducing \texttt{saha\_ncomps} below 3 is not viable as it degrades light-curve quality significantly, while increasing it adds cost with no accuracy benefit, so the per-call work is already at its minimum. For zones deep in the optically thick interior, where temperature and density change by $\lesssim0.1\%$ per iteration, caching the Saha result and recomputing only when conditions exceed a threshold could skip ${\sim}50$--$70\%$ of the N-R solves, since the outer ${\sim}30$--$50$ zones (where the photosphere resides and conditions change rapidly) would still require full evaluation. Given that the EOS is a significant fraction of the per-timestep cost, this could yield a ${\sim}10$--$20\%$ overall speedup at $N=100$, with a larger fractional gain at higher $N$ where the Saha cost per call is unchanged but the number of interior zones grows. Whether this is worth the added code complexity is left for future work.

\textit{Thread-level parallelism.} The zone-by-zone equation-of-state, opacity, and Ni deposition loops are embarrassingly parallel in principle. However, at 100 zones the per-model runtime is already $<2$~s, so OpenMP thread management overhead dominates any potential gain; in our tests an OpenMP-enabled build ran slightly \emph{slower} than the serial baseline. For large model grids, trivial parallelism by launching independent SNEC instances is both simpler and more effective. Thread-level parallelism could still be relevant for single high-zone-count runs ($N \gtrsim 1000$), where the per-zone work is large enough to amortize the overhead.

\section{Reproducibility}
The SuperSNEC source code, analysis scripts, and all data needed to reproduce the tables and figures in this paper are publicly available at \url{https://github.com/cfremling/SuperSNEC}. The repository includes the complete Fortran source, a ready-to-use parameter file, and progenitor stellar profiles. Building the code requires only a Fortran compiler (e.g., \texttt{gfortran}) and a LAPACK library; compilation is performed with a single \texttt{make} command.

All parameter sweeps, tables, and figures presented in this work can be regenerated from scratch by running \texttt{python Analysis/\-scripts/\-rebuild\_all.py} from the repository root. This master script executes each parameter sweep (compiling and running SuperSNEC as needed), collects the results, and produces the \LaTeX{} tables and figures automatically. Individual sweep scripts in \texttt{Analysis/\-scripts/} can also be run independently; each accepts a \texttt{-{}-tables-only} flag to regenerate tables from cached results without re-running the simulations. The analysis scripts require Python~$\geq$~3.10 with NumPy, SciPy, and Matplotlib.

\begin{acknowledgments}
We acknowledge the excellent work by the original authors of SNEC that has made this work possible. Large language models (Claude, Anthropic; GPT/Codex, OpenAI) were essential tools for code optimization, systematic validation, and manuscript preparation. Christoffer Fremling is supported by NSF AST-2407588.
\end{acknowledgments}

\software{SNEC \citep{Morozova2015}, MESA \citep{Paxton2011, Paxton2013, Paxton2015, Paxton2018, Paxton2019}, GNU Fortran (GCC), gnuplot \citep{gnuplot}, NumPy \citep{Harris2020}, Matplotlib \citep{Hunter2007}, Python, Claude Opus 4.6 \citep{ClaudeOpus46}, GPT-5 \citep{GPT5SystemCard}, Codex GPT-5.3 \citep{GPT53Codex}}

\appendix

\section{Speed-Optimization Details}
\label{sec:speed_details}

This appendix presents the parameter sweep results that underlie the cumulative speedup of SuperSNEC summarized in Sect.~\ref{sec:cumulative_speedup} and Table~\ref{tab:cumulative_speedup}. All sweeps use the ``He4'' stripped-envelope helium-star progenitor from \citet{Morozova2015} ($15\,\Msun$ ZAMS, binary-stripped; Sect.~\ref{sec:validation}) exploded as a thermal bomb with $E_k = 10^{51}$~erg. Unless otherwise noted, runs use the \texttt{adaptive\_runtime} grid mode at $N = 100$ zones and are compared against the same 1000-zone SNEC reference run described in Sect.~\ref{sec:validation} (runtime $671$~s, $\mathrm{EPSTOL} = 10^{-7}$, fixed \texttt{GridPattern.dat} grid). RMS magnitude residuals are evaluated in the phase bins 0--5~d, 5--30~d, and $>30$~d.

\subsection{Solver tolerance}
\label{sec:solver_sweep}

We performed isolated one-parameter scans of $\mathrm{EPSTOL}$ and $\mathrm{ITMAX}$ in \texttt{adaptive\_\-runtime} mode at $N = 100$ zones. Table~\ref{tab:solver_itmax_only} shows the $\mathrm{ITMAX}$-only scan at fixed $\mathrm{EPSTOL}=10^{-7}$; Table~\ref{tab:solver_epstol_only} shows the $\mathrm{EPSTOL}$ scan at fixed $\mathrm{ITMAX}(100,300)$.

Reducing $\mathrm{ITMAX}$ is counterproductive: a reduction from $(100,300)$ to $(5,15)$ increases the runtime by ${\sim}60\%$ due to the accumulation of scratched timesteps (timesteps where the solver fails to converge within $\mathrm{ITMAX}$ iterations, forcing the code to discard the step, halve the timestep, and retry) while the RMS residual remains unchanged. At $\mathrm{ITMAX}=(1,3)$ the simulation fails entirely. Moreover, the $\mathrm{ITMAX}$ scan was performed at the default $\mathrm{EPSTOL}=10^{-7}$; at a relaxed $\mathrm{EPSTOL}=10^{-4}$ the solver converges in fewer iterations and $\mathrm{ITMAX}$ is never reached (zero scratched steps), making it effectively inactive as a limiter. As such, we see no reason to raise $\mathrm{ITMAX}$ above its default, which could otherwise be motivated if scratched steps were a significant runtime contributor at the adopted $\mathrm{EPSTOL}$.

The dominant runtime lever among these two parameters is therefore $\mathrm{EPSTOL}$. Relaxing $\mathrm{EPSTOL}$ from $10^{-7}$ to $10^{-4}$ reduces the runtime from $4.3$~s to $1.6$~s with negligible impact on accuracy: the RMS change is $<0.005$~mag in all phase bins. Beyond $10^{-4}$, accuracy degrades progressively: at $\mathrm{EPSTOL} = 10^{-3}$ the 5--30~d RMS rises to $0.11$~mag, and at $\mathrm{EPSTOL} = 2 \times 10^{-3}$ it reaches $0.13$~mag. We adopt $\mathrm{EPSTOL}=10^{-4}$ with $\mathrm{ITMAX}(100,300)$ as the solver baseline for all subsequent parameter scans. This value sits at the accuracy knee: the RMS residual is flat from $10^{-7}$ to $10^{-4}$ but degrades sharply beyond, making $10^{-4}$ the most aggressive tolerance that does not sacrifice accuracy.

\subsection{Quadrature resolution}
\label{sec:ni_quad}

The gamma-ray deposition calculation in \texttt{nickel.F90} uses an $n_\mathrm{quad} \times n_\mathrm{quad}$ quadrature (\texttt{Ni\_quad\_\-npoints}) for the angular and radial integration of the optical depth to gamma rays. The original SNEC default of $n_\mathrm{quad} = 150$ ($22{,}500$ evaluations per zone) was chosen for 1000-zone models. We performed a convergence study at $N = 100$ zones, scanning $n_\mathrm{quad}$ from 30 to 150 (Table~\ref{tab:quad_convergence}). The RMS residual is essentially flat for $n_\mathrm{quad} \ge 70$ ($\Delta m_\mathrm{all} \approx 0.024$--$0.027$~mag), indicating that the quadrature is heavily over-resolved at 100 zones. We adopt $n_\mathrm{quad} = 70$ as part of our baseline settings, which cuts the $N=100$ runtime from $2.3$~s to $1.6$~s relative to $n_\mathrm{quad}=150$ with no measurable accuracy loss. Reducing to $n_\mathrm{quad} = 50$ introduces a modest late-time degradation ($\Delta m_\mathrm{all} = 0.033$ vs $0.026$). The quadrature flatness above $n_\mathrm{quad} = 70$ is a consequence of the coarse spatial grid: at 100 zones the optical-depth structure is smooth enough that fine angular/radial sampling adds no information. Users running at higher $N$ (e.g., 500--1000 zones) should increase $n_\mathrm{quad}$ toward the SNEC default of 150 to ensure convergence of the finer spatial structure.

\subsection{$^{56}$Ni deposition cadence}
\label{sec:ni_cadence_sweep}

The $^{56}$Ni heating calculation is the most expensive per-call operation in SNEC. In SuperSNEC its update cadence is controlled by three parameters: $\Delta t_{\mathrm{Ni,min}}$ (\texttt{Ni\_period}), $\Delta t_{\mathrm{Ni,max}}$ (\texttt{Ni\_period\_\-max}), and $f_{\mathrm{Ni}}$ (\texttt{Ni\_fractional\_\-change}). In \texttt{adaptive\_\-runtime} mode, the Ni update cadence operates in two distinct regimes. During the remapping phase ($q_\mathrm{photo} > q_\mathrm{stop}$; $t \lesssim 34$~d for our test model), each grid remap forces a Ni recalculation (Sect.~\ref{sec:ni_remap_interaction}), so the effective update rate is set by the remap interval (${\sim}1$~d) regardless of $f_{\mathrm{Ni}}$. After remapping is permanently disabled ($q_\mathrm{photo} \le q_\mathrm{stop}$; Sect.~\ref{sec:remap_scheduling}), the cadence timer with $f_{\mathrm{Ni}}$ becomes the sole controller of Ni update frequency. The choice of $f_{\mathrm{Ni}}$ is therefore largely irrelevant during the remapping phase but directly controls late-time accuracy.

Table~\ref{tab:ni_cadence_comparison} quantifies this by running matched Ni cadence settings across both grid modes. In \texttt{adaptive\_\-runtime} mode, the early- and mid-phase RMS (0--30~d) is insensitive to $f_{\mathrm{Ni}}$ because remap-forced updates dominate; but the late-phase RMS ($>30$~d) degrades at aggressive settings, rising from $0.004$~mag at $f_{\mathrm{Ni}} = 0.10$ to $0.026$~mag at $f_{\mathrm{Ni}} = 0.70$. A moderate $f_{\mathrm{Ni}} = 0.20$ (50~Ni evaluations) yields $\Delta m_\mathrm{all} = 0.022$~mag, retaining most of the speed benefit while keeping the late-time residual well controlled. In \texttt{legacy\_\-pattern} mode, no remap-forced updates occur, and aggressive $f_{\mathrm{Ni}}$ degrades accuracy across all phases: at $f_{\mathrm{Ni}} = 0.70$ the RMS reaches $0.19$~mag with only 14~Ni evaluations. The legacy grid requires $f_{\mathrm{Ni}} \leq 0.10$ ($94$--$101$ evals) to keep the 5--30~d bin below $0.1$~mag.

Table~\ref{tab:legacy_ni_cadence} shows a two-parameter \texttt{legacy-pattern} Ni-cadence sweep, varying $f_{\mathrm{Ni}}$ and $\Delta t_{\mathrm{Ni,max}}$ jointly at fixed $\Delta t_{\mathrm{Ni,min}} = 5\times10^4$~s. At conservative cadence ($f_{\mathrm{Ni}} \leq 0.10$), the RMS is insensitive to $\Delta t_{\mathrm{Ni,max}}$ and stays below 0.1~mag. At aggressive cadence ($f_{\mathrm{Ni}} \geq 0.50$), accuracy degrades once $\Delta t_{\mathrm{Ni,max}}$ is large enough to reduce Ni evaluations below ${\sim}60$, with the 5--30~d RMS exceeding 0.1~mag. Increasing $\Delta t_{\mathrm{Ni,max}}$ reduces Ni evaluations and runtime but does not compensate for aggressive $f_{\mathrm{Ni}}$. In the \texttt{legacy-pattern} fixed runtime grid mode, the $^{56}$Ni updates need to be kept frequent either through low $f_{\mathrm{Ni}}$ or low $\Delta t_{\mathrm{Ni,max}}$ (on the order of $0.5$~days).

We adopt $f_{\mathrm{Ni}} = 0.20$ with $\Delta t_{\mathrm{Ni,min}} = 5\times10^4$~s and $\Delta t_{\mathrm{Ni,max}} = 5.456\times10^5$~s as the Ni-cadence baseline. This balances speed (roughly half the Ni evaluations of the most conservative settings) with late-time accuracy: the $>30$~d RMS remains below 0.006~mag rather than rising to 0.026~mag at $f_{\mathrm{Ni}} = 0.70$. In \texttt{legacy\_\-pattern} mode a more conservative $f_{\mathrm{Ni}} \leq 0.10$ is required.

\subsection{Remap cadence}
\label{sec:grid_cadence_sweep}
Table~\ref{tab:remap_scheduling} shows the effect of the remap interval $\Delta t_{\mathrm{remap}}$ on accuracy and runtime at 100 zones evolved to 300~d. With $q_\mathrm{stop} = 0.50$ (Sect.~\ref{sec:remap_scheduling}), remapping is permanently disabled once the photosphere recedes below $q = 0.50$ ($t \approx 34$~d for our test model), so all interpolation diffusion is confined to the early remapping phase. Very frequent updates ($\leq0.5$~d) increase the remap and Ni evaluation counts without significantly improving peak-phase accuracy, because the grid target changes negligibly between consecutive checks. Increasing the interval to 2--3~d reduces these counts but sharply degrades $\Delta m_{5\text{--}30}$ as the grid lags the rapid early photosphere recession. The late-time residual ($\Delta m_{>30}$) shows a mild opposite trend. It is slightly lower for infrequent remaps, but this effect is modest compared to the peak-phase degradation. We adopt $\Delta t_{\mathrm{remap}} = 1.0$~d as our baseline, which minimizes $\Delta m_\mathrm{all}$. The total number of remaps is determined by $\Delta t_{\mathrm{remap}}$ and the time for the photosphere to recede to $q_\mathrm{stop}$, which depends on the progenitor and explosion parameters.

We note that for $\Delta t_{\mathrm{remap}} = 1.0$~d there is a modest increase in RMS at very late times ($\Delta m_{>200}$). This is hinting at a real limitation of very low zone count models. Studies that focus on late-time evolution should consider increasing the zone count to $\geq200$, which is where we find that this increasing RMS trend flattens when compared to 1000-zone models. 

\subsection{Maximum timestep}
\label{sec:dtmax_sweep}

The $\Delta t_{\mathrm{max}}$ parameter (\texttt{dtmax}) sets a hard ceiling on the hydrodynamic timestep. The actual timestep is the minimum of $\Delta t_{\mathrm{max}}$ and the CFL condition $\Delta t_\mathrm{CFL} = 0.95 \times \min_i \Delta r_i / (|v_i| + c_{s,i})$, with a growth limiter that restricts step-to-step increases to 2.5\%.

We swept $\Delta t_{\mathrm{max}}$ from $10^3$ to $10^6$~s at 100 zones with $t_\mathrm{end} = 120$~d. Above $\Delta t_{\mathrm{max}} \geq 2\times10^5$~s the CFL condition becomes the sole constraint and further increases have no effect on timestep count, runtime, or accuracy. The original SNEC default of $10^4$~s is unnecessarily restrictive. We adopt $\Delta t_{\mathrm{max}} = 10^5$~s, which sits in the flat regime and reduces the total number of timesteps by ${\sim}20\%$ relative to the SNEC default. We choose $10^5$~s over $2\times10^5$~s (where the CFL fully dominates) as a conservative choice: at $10^5$~s the ceiling is binding only during the first few days when velocities are highest, and by $t \gtrsim 10$~d the CFL condition is already more restrictive. Any value above $10^5$~s produces identical results, so the choice is simply the lowest value in the flat regime, providing a safety margin against edge-case configurations (e.g., lower-energy explosions with slower ejecta) where the CFL transition may shift.

\subsection{Output cadence and output mode}
\label{sec:output_cadence_speed}

The output cadence (Sect.~\ref{sec:cadence}) and output mode (Sect.~\ref{sec:output_mode}) affect runtime through file writing rather than computation. We benchmarked both effects at the working baseline settings ($N = 100$).

The choice of output cadence has negligible impact on runtime for reasonable settings. With full output (\texttt{output\_mode\,=\,0}), varying the scalar cadence across a factor of ${\sim}5$ (from ${\sim}35$ to ${\sim}184$ dumps) changes the wall-clock time by only ${\sim}3\%$ ($1.59$--$1.64$~s). Profile cadence has a somewhat larger effect: increasing the number of profile dumps from the baseline (${\sim}35$) to match the scalar cadence (${\sim}140$ dumps, or ${\sim}2$ per day on average) adds ${\sim}30\%$ overhead due to the 29 appended \texttt{.xg} files, a setting useful for detailed profile analysis but unnecessary for light-curve fitting alone. With minimal output (\texttt{output\_mode\,=\,1}), where profile files are suppressed entirely, neither scalar nor profile cadence has any measurable effect on runtime (${\sim}1.3$~s across all settings tested). The output cadence parameters can therefore be chosen purely for light-curve sampling quality without runtime concern.

The minimal output mode itself provides a ${\sim}1.2\times$ speedup over full output at both the baseline and fast configurations (Table~\ref{tab:cumulative_speedup}), consistent with the profile-writing overhead measured above. For single-model analysis and validation, full output (mode~0) is recommended to retain radial profiles for diagnostic inspection. For production model-grid runs ($\gtrsim10^4$ models), minimal output (mode~1) is recommended to limit disk usage (${\sim}50$~KB vs ${\sim}6$~MB per model) while retaining all data needed for light-curve fitting.

\section{Configuration Parameters}
\label{sec:parameters}

Table~\ref{tab:parameters} lists all user-facing configuration parameters in SNEC and SuperSNEC. Parameters marked with $\dagger$ are new additions in SuperSNEC; all others are inherited from SNEC~1.01. In the original SNEC, $\mathrm{EPSTOL}$ and $\mathrm{ITMAX}$ are compiled constants that require recompilation to change; SuperSNEC exposes them as runtime parameters (\texttt{epstol\_hydro}, \texttt{epstol\_rad}, \texttt{itmax\_hydro}, \texttt{itmax\_rad}). The ``Working value'' column shows the optimized baseline used throughout this work (100-zone \texttt{adaptive\_\-runtime} mode); dashes indicate parameters not active in the baseline configuration.


\begin{deluxetable*}{cccccc}
\tablecaption{ITMAX-only scan at fixed \texttt{EPSTOL}$=10^{-7}$ (adaptive-runtime setup).\label{tab:solver_itmax_only}}
\tablehead{
\colhead{$\mathrm{ITMAX}_{\mathrm{hyd}}$} &
\colhead{$\mathrm{ITMAX}_{\mathrm{rad}}$} &
\colhead{Runtime (s)} &
\colhead{Scratch} &
\colhead{RMS 5--30 d} &
\colhead{RMS $>30$ d}
}
\startdata
100 & 300 & 4.29 & 4 & 0.0265 & 0.0056 \\
80 & 240 & 4.79 & 7 & 0.0259 & 0.0072 \\
60 & 200 & 4.91 & 12 & 0.0230 & 0.0055 \\
40 & 150 & 4.96 & 20 & 0.0236 & 0.0059 \\
30 & 120 & 5.22 & 26 & 0.0233 & 0.0075 \\
20 & 90 & 5.40 & 40 & 0.0232 & 0.0054 \\
15 & 60 & 6.05 & 66 & 0.0245 & 0.0061 \\
10 & 45 & 5.86 & 102 & 0.0242 & 0.0072 \\
8 & 30 & 6.28 & 174 & 0.0231 & 0.0061 \\
6 & 20 & 6.67 & 275 & 0.0236 & 0.0071 \\
5 & 15 & 7.04 & 382 & 0.0228 & 0.0041 \\
4 & 12 & 7.41 & 489 & 0.0240 & 0.0058 \\
3 & 9 & 8.08 & 687 & 0.0234 & 0.0058 \\
2 & 6 & 9.55 & 1142 & 0.0230 & 0.0060 \\
\enddata
\tablecomments{Lowering \texttt{ITMAX} increases retry cost and does not produce reliable runtime gains in this fixed-\texttt{EPSTOL} test.}
\end{deluxetable*}

\begin{deluxetable*}{ccccc}
\tablecaption{EPSTOL sweep at fixed \texttt{ITMAX}$(\mathrm{hyd},\mathrm{rad})=(100,300)$ (adaptive-runtime setup).\label{tab:solver_epstol_only}}
\tablehead{
\colhead{$\mathrm{EPSTOL}$} &
\colhead{Runtime (s)} &
\colhead{Scratch} &
\colhead{RMS 5--30 d} &
\colhead{RMS $>30$ d}
}
\startdata
1.0d-7 & 4.29 & 4 & 0.0265 & 0.0056 \\
3.0d-7 & 3.49 & 1 & 0.0242 & 0.0051 \\
1.0d-6 & 2.83 & 0 & 0.0253 & 0.0056 \\
2.0d-6 & 2.51 & 0 & 0.0255 & 0.0051 \\
5.0d-6 & 2.24 & 0 & 0.0249 & 0.0052 \\
1.0d-5 & 2.03 & 0 & 0.0252 & 0.0061 \\
2.0d-5 & 1.99 & 0 & 0.0257 & 0.0050 \\
5.0d-5 & 1.66 & 0 & 0.0259 & 0.0064 \\
1.0d-4 & 1.59 & 0 & 0.0262 & 0.0048 \\
2.0d-4 & 1.57 & 0 & 0.0297 & 0.0071 \\
5.0d-4 & 1.44 & 0 & 0.0585 & 0.0140 \\
1.0d-3 & 1.27 & 0 & 0.1056 & 0.0282 \\
2.0d-3 & 1.27 & 0 & 0.1347 & 0.0321 \\
\enddata
\tablecomments{Fixed \texttt{ITMAX}$(\mathrm{hyd},\mathrm{rad})=(100,300)$. The dominant runtime lever is \texttt{EPSTOL}; values around $10^{-4}$ are fast, while $\gtrsim2\times10^{-4}$ gradually degrades quality (specially at $5-30$~d) or fails.}
\end{deluxetable*}

\begin{deluxetable*}{rrrr}
\setlength{\tabcolsep}{12pt}
\tablecaption{Quadrature convergence at $N = 100$ zones. $n_\mathrm{quad}$ is the number of integration points per dimension in the gamma-ray deposition calculation. All rows use the adaptive baseline (Sect.~\ref{sec:baseline}) except for the varied $n_\mathrm{quad}$. RMS is measured against a 1000-zone SNEC reference. \label{tab:quad_convergence}}
\tablehead{
\colhead{\hfill$n_\mathrm{quad}$} & \colhead{$t$ (s)} & \colhead{$\Delta m_{5\text{--}30}$} & \colhead{$\Delta m_\mathrm{all}$}
}
\startdata
\cutinhead{SuperSNEC ($N = 100$)}
  30 &   1.5 & 0.040 & 0.069 \\
  50 &   1.6 & 0.026 & 0.033 \\
  70 &   1.6 & 0.026 & 0.026 \\
  90 &   1.7 & 0.029 & 0.025 \\
 110 &   1.9 & 0.032 & 0.027 \\
 130 &   2.1 & 0.032 & 0.027 \\
 150 &   2.3 & 0.033 & 0.027 \\
\enddata
\tablecomments{$\Delta m_{5\text{--}30}$ and $\Delta m_\mathrm{all}$ are RMS magnitude residuals in the 5--30\,d and $>$0.5\,d windows. The RMS is essentially flat for $n_\mathrm{quad} \ge 50$.}
\end{deluxetable*}

\begin{deluxetable*}{lccccccc}
\tablecaption{Ni-cadence tolerance comparison across grid modes at matched settings ($\Delta t_{\mathrm{min}} = 5\times10^4$~s, $\Delta t_{\mathrm{max}} = 5.456\times10^5$~s, 100 zones). Each row shows a paired \texttt{adaptive\_runtime} and \texttt{legacy\_pattern} run at the same $f_{\mathrm{Ni}}$.\label{tab:ni_cadence_comparison}}
\tablehead{
\colhead{$f_{\mathrm{Ni}}$} &
\colhead{Grid mode} &
\colhead{Runtime (s)} &
\colhead{Ni evals} &
\colhead{RMS 0--5\,d} &
\colhead{RMS 5--30\,d} &
\colhead{RMS $>30$\,d} &
\colhead{RMS all}
}
\startdata
0.05 & \texttt{adaptive} & 1.83 & 75 & 0.0462 & 0.0266 & 0.0043 & 0.0222 \\
0.05 & \texttt{legacy} & 1.77 & 101 & 0.1765 & 0.0866 & 0.0389 & 0.0812 \\
\hline
0.10 & \texttt{adaptive} & 1.71 & 74 & 0.0469 & 0.0241 & 0.0034 & 0.0210 \\
0.10 & \texttt{legacy} & 1.72 & 94 & 0.1767 & 0.0869 & 0.0392 & 0.0814 \\
\hline
0.20 & \texttt{adaptive} & 1.55 & 50 & 0.0473 & 0.0262 & 0.0048 & 0.0223 \\
0.20 & \texttt{legacy} & 1.44 & 46 & 0.1779 & 0.1105 & 0.0511 & 0.0959 \\
\hline
0.70 & \texttt{adaptive} & 1.50 & 34 & 0.0473 & 0.0262 & 0.0270 & 0.0289 \\
0.70 & \texttt{legacy} & 1.27 & 14 & 0.2022 & 0.2460 & 0.1036 & 0.1866 \\
\hline
1.00 & \texttt{adaptive} & 1.47 & 33 & 0.0473 & 0.0262 & 0.0404 & 0.0357 \\
1.00 & \texttt{legacy} & 1.27 & 10 & 0.2107 & 0.3472 & 0.1389 & 0.2555 \\
\enddata
\tablecomments{In \texttt{adaptive\_runtime} mode, grid remaps reset the Ni cadence timer, producing supplementary Ni evaluations that are included in the Ni evals count. The \texttt{legacy\_pattern} mode has no remap-forced updates. This mechanism allows \texttt{adaptive\_runtime} to tolerate aggressive $f_{\mathrm{Ni}}$ values that would degrade accuracy in \texttt{legacy\_pattern}.}
\end{deluxetable*}

\begin{deluxetable*}{ccccccccccc}
\tablecaption{Effect of remap cadence on accuracy and runtime evolved to 300\,d. All runs use the working baseline from Sect.~\ref{sec:baseline}.\label{tab:remap_scheduling}}
\tablehead{
\colhead{Interval} &
\colhead{Remaps} &
\colhead{Ni evals} &
\colhead{Runtime} &
\colhead{$\Delta m_{0\text{--}5}$} &
\colhead{$\Delta m_{5\text{--}30}$} &
\colhead{$\Delta m_{30\text{--}100}$} &
\colhead{$\Delta m_{100\text{--}200}$} &
\colhead{$\Delta m_{200\text{--}300}$} &
\colhead{$\Delta m_{>30}$} &
\colhead{$\Delta m_\mathrm{all}$}
\\
\colhead{(d)} &
\colhead{} &
\colhead{} &
\colhead{(s)} &
\colhead{(mag)} &
\colhead{(mag)} &
\colhead{(mag)} &
\colhead{(mag)} &
\colhead{(mag)} &
\colhead{(mag)} &
\colhead{(mag)}
}
\startdata
\cutinhead{$N = 100$}
0.50 & 51 & 293 & 3.45 & 0.0413 & 0.0246 & 0.0123 & 0.0325 & 0.0617 & 0.0429 & 0.0417 \\
1.0 & 28 & 270 & 3.41 & 0.0478 & 0.0260 & 0.0127 & 0.0302 & 0.0556 & 0.0390 & 0.0383 \\
2.0 & 14 & 254 & 4.40 & 0.0560 & 0.0759 & 0.0157 & 0.0256 & 0.0497 & 0.0350 & 0.0404 \\
3.0 & 9 & 138 & 3.94 & 0.0608 & 0.1001 & 0.0196 & 0.0161 & 0.0359 & 0.0260 & 0.0387 \\
\enddata
\tablecomments{All runs use $q_\mathrm{stop} = 0.50$, so remapping is disabled after $t \approx 34$\,d.  Intervals $\geq 2$\,d degrade $\Delta m_{5\text{--}30}$ as the grid lags the photosphere recession.  The 1.0\,d baseline minimizes $\Delta m_\mathrm{all}$.}
\end{deluxetable*}

\begin{deluxetable*}{cccccccc}
\tablecaption{Two-parameter $^{56}$Ni-cadence sweep in \texttt{legacy\_pattern} mode at 100 zones. \texttt{Ni\_fractional\_change} ($f_{\mathrm{Ni}}$) and \texttt{Ni\_period\_max} ($\Delta t_{\mathrm{max}}$) are varied jointly; \texttt{Ni\_period}$=5\times10^4$~s is fixed. Solver baseline: \texttt{EPSTOL}$=10^{-4}$, \texttt{ITMAX}$(100,300)$.\label{tab:legacy_ni_cadence}}
\tablehead{
\colhead{$f_{\mathrm{Ni}}$} &
\colhead{$\Delta t_{\mathrm{max}}$ (s)} &
\colhead{Runtime (s)} &
\colhead{Ni evals} &
\colhead{RMS 0--5\,d} &
\colhead{RMS 5--30\,d} &
\colhead{RMS $>30$\,d} &
\colhead{RMS all}
}
\startdata
0.05 & $5\times10^4$ & 2.25 & 101 & 0.1765 & 0.0866 & 0.0389 & 0.0812 \\
0.05 & $1.73\times10^5$ & 1.86 & 101 & 0.1765 & 0.0866 & 0.0389 & 0.0812 \\
0.05 & $3.46\times10^5$ & 1.91 & 101 & 0.1765 & 0.0866 & 0.0389 & 0.0812 \\
0.05 & $5.46\times10^5$ & 1.76 & 101 & 0.1765 & 0.0866 & 0.0389 & 0.0812 \\
0.05 & $8.64\times10^5$ & 1.76 & 101 & 0.1765 & 0.0866 & 0.0389 & 0.0812 \\
0.05 & $1.73\times10^6$ & 1.58 & 70 & 0.1765 & 0.0892 & 0.0433 & 0.0834 \\
\hline
0.10 & $5\times10^4$ & 1.75 & 101 & 0.1765 & 0.0866 & 0.0389 & 0.0812 \\
0.10 & $1.73\times10^5$ & 1.76 & 101 & 0.1765 & 0.0866 & 0.0389 & 0.0812 \\
0.10 & $3.46\times10^5$ & 1.72 & 101 & 0.1765 & 0.0866 & 0.0389 & 0.0812 \\
0.10 & $5.46\times10^5$ & 1.72 & 94 & 0.1767 & 0.0869 & 0.0392 & 0.0814 \\
0.10 & $8.64\times10^5$ & 1.49 & 60 & 0.1770 & 0.1018 & 0.0456 & 0.0901 \\
0.10 & $1.73\times10^6$ & 1.38 & 34 & 0.1772 & 0.1142 & 0.0586 & 0.0997 \\
\hline
0.20 & $5\times10^4$ & 1.72 & 101 & 0.1765 & 0.0866 & 0.0389 & 0.0812 \\
0.20 & $1.73\times10^5$ & 1.75 & 101 & 0.1765 & 0.0866 & 0.0389 & 0.0812 \\
0.20 & $3.46\times10^5$ & 1.59 & 77 & 0.1771 & 0.0937 & 0.0422 & 0.0854 \\
0.20 & $5.46\times10^5$ & 1.44 & 46 & 0.1779 & 0.1105 & 0.0511 & 0.0959 \\
0.20 & $8.64\times10^5$ & 1.38 & 30 & 0.1810 & 0.1416 & 0.0588 & 0.1144 \\
0.20 & $1.73\times10^6$ & 1.32 & 18 & 0.1818 & 0.1612 & 0.0892 & 0.1337 \\
\hline
0.50 & $5\times10^4$ & 1.74 & 101 & 0.1765 & 0.0866 & 0.0389 & 0.0812 \\
0.50 & $1.73\times10^5$ & 1.50 & 60 & 0.1770 & 0.1018 & 0.0456 & 0.0901 \\
0.50 & $3.46\times10^5$ & 1.33 & 30 & 0.1810 & 0.1416 & 0.0588 & 0.1144 \\
0.50 & $5.46\times10^5$ & 1.27 & 19 & 0.1888 & 0.1888 & 0.0809 & 0.1469 \\
0.50 & $8.64\times10^5$ & 1.27 & 13 & 0.2107 & 0.2775 & 0.1157 & 0.2087 \\
0.50 & $1.73\times10^6$ & 1.27 & 8 & 0.2107 & 0.3196 & 0.1771 & 0.2512 \\
\hline
0.70 & $5\times10^4$ & 1.81 & 101 & 0.1765 & 0.0866 & 0.0389 & 0.0812 \\
0.70 & $1.73\times10^5$ & 1.43 & 42 & 0.1782 & 0.1177 & 0.0515 & 0.0997 \\
0.70 & $3.46\times10^5$ & 1.32 & 22 & 0.1867 & 0.1729 & 0.0749 & 0.1363 \\
0.70 & $5.46\times10^5$ & 1.26 & 14 & 0.2022 & 0.2460 & 0.1036 & 0.1866 \\
0.70 & $8.64\times10^5$ & 1.27 & 10 & 0.2107 & 0.3817 & 0.1519 & 0.2793 \\
0.70 & $1.73\times10^6$ & 1.28 & 7 & 0.2107 & 0.4331 & 0.2400 & 0.3358 \\
\hline
1.00 & $5\times10^4$ & 1.75 & 101 & 0.1765 & 0.0866 & 0.0389 & 0.0812 \\
1.00 & $1.73\times10^5$ & 1.38 & 30 & 0.1810 & 0.1416 & 0.0588 & 0.1144 \\
1.00 & $3.46\times10^5$ & 1.32 & 16 & 0.1966 & 0.2266 & 0.0975 & 0.1734 \\
1.00 & $5.46\times10^5$ & 1.27 & 10 & 0.2107 & 0.3472 & 0.1389 & 0.2555 \\
1.00 & $8.64\times10^5$ & 1.26 & 7 & 0.2107 & 0.5350 & 0.2056 & 0.3849 \\
1.00 & $1.73\times10^6$ & 1.26 & 5 & 0.2107 & 0.5772 & 0.3290 & 0.4476 \\
\enddata
\tablecomments{36 of 36 cases complete successfully. The fixed grid has no remap-forced Ni recalculations. Increasing $\Delta t_{\mathrm{max}}$ reduces Ni evaluations and runtime, but at aggressive $f_{\mathrm{Ni}}$ the late-time RMS degrades sharply. Conservative settings ($f_{\mathrm{Ni}} \leq 0.10$) are required for sub-0.1\,mag accuracy.}
\end{deluxetable*}


\begin{deluxetable*}{llccc}
\tabletypesize{\scriptsize}
\tablecaption{SNEC and SuperSNEC configuration parameters. A $\dagger$ marks parameters introduced by SuperSNEC. The working values correspond to the 100-zone \texttt{adaptive\_\-runtime} baseline used throughout this paper.\label{tab:parameters}}
\tablehead{
\colhead{Parameter} &
\colhead{Description} &
\colhead{Default} &
\colhead{Working value} &
\colhead{Units}
}
\startdata
\multicolumn{5}{c}{\textit{Grid}} \\
\hline
\texttt{imax} & Number of radial zones & 1000 & 100 & \nodata \\
\texttt{grid\_mode}$^{\dagger}$ & Grid initialization mode & \texttt{adaptive\_runtime} & \texttt{adaptive\_runtime} & \nodata \\
\texttt{grid\_pattern\_file}$^{\dagger}$ & Fixed-grid pattern file (\texttt{legacy\_pattern} only) & \texttt{GridPattern.dat} & --- & \nodata \\
\texttt{grid\_surface\_alpha}$^{\dagger}$ & Surface concentration exponent $\alpha_{\mathrm{surf}}$ & 5.0 & 7.0 & \nodata \\
\texttt{grid\_relax\_days}$^{\dagger}$ & Grid relaxation timescale $t_{\mathrm{relax}}$ & 2.0 & 5.0 & d \\
\texttt{grid\_update\_interval\_days}$^{\dagger}$ & Adaptive remap cadence $\Delta t_{\mathrm{remap}}$ & 0.0 & 1.0 & d \\
\texttt{grid\_min\_cell\_frac}$^{\dagger}$ & Min.\ outer-cell width / uniform-cell width & $10^{-4}$ & $10^{-4}$ & \nodata \\
\texttt{grid\_remap\_qphoto\_stop}$^{\dagger}$ & Photosphere $q$ threshold to stop remapping & 0.50 & 0.50 & \nodata \\
\texttt{mass\_excision} & Excise inner mass & yes & yes & \nodata \\
\texttt{mass\_excised} & Excised mass & 1.4 & 1.4 & $M_{\odot}$ \\
\hline
\multicolumn{5}{c}{\textit{Explosion}} \\
\hline
\texttt{initial\_data} & Explosion mechanism & \texttt{Thermal\_Bomb} & \texttt{Thermal\_Bomb} & \nodata \\
\texttt{final\_energy} & Asymptotic kinetic energy & $10^{51}$ & $10^{51}$ & erg \\
\texttt{bomb\_tstart} & Energy injection start & 0.0 & 0.0 & s \\
\texttt{bomb\_tend} & Energy injection end & 0.1 & 0.1 & s \\
\texttt{bomb\_mass\_spread} & Energy deposition mass range & 0.1 & 0.1 & $M_{\odot}$ \\
\hline
\multicolumn{5}{c}{\textit{$^{56}$Ni heating and mixing}} \\
\hline
\texttt{Ni\_switch} & Enable $^{56}$Ni heating & 1 (on) & 1 (on) & \nodata \\
\texttt{Ni\_mass} & Initial $^{56}$Ni mass & 0.03 & 0.03 & $M_{\odot}$ \\
\texttt{Ni\_by\_hand} & Manual Ni placement mode & 1 & 1 & \nodata \\
\texttt{Ni\_period} & Minimum Ni update interval $\Delta t_{\mathrm{Ni,min}}$ & $5\times10^{4}$ & $5\times10^{4}$ & s \\
\texttt{Ni\_period\_max}$^{\dagger}$ & Maximum Ni update interval $\Delta t_{\mathrm{Ni,max}}$ & $10\times\Delta t_{\mathrm{Ni,min}}$ & $5.456\times10^{5}$ & s \\
\texttt{Ni\_fractional\_change}$^{\dagger}$ & Fractional change trigger $f_{\mathrm{Ni}}$ & 0.25 & 0.20 & \nodata \\
\texttt{Ni\_mix\_fraction}$^{\dagger}$ & Ni mixing extent (ejecta mass fraction) & --- & 0.31 & \nodata \\
\texttt{Ni\_mix\_kernel}$^{\dagger}$ & Mixing kernel (1\,=\,box, 2\,=\,exponential, 3\,=\,two-component) & 2 & 1 & \nodata \\
\texttt{Ni\_mix\_component2\_fraction}$^{\dagger}$ & Outer-shell Ni fraction (kernel\,3 only) & 0.0 & --- & \nodata \\
\texttt{Ni\_mix\_component2\_extent}$^{\dagger}$ & Outer-shell extent $q_{\mathrm{mix,2}}$ (kernel\,3 only) & $q_{\mathrm{mix}}$ & --- & \nodata \\
\texttt{Ni\_quad\_npoints}$^{\dagger}$ & Quadrature points per dimension (Sect.~\ref{sec:ni_quad}) & 150 & 70 & \nodata \\
\texttt{ni\_raytrace\_opt}$^{\dagger}$ & Gamma-ray tracing optimization (0\,=\,off, 1\,=\,on) & 1 & 1 & \nodata \\
\texttt{ni\_ray\_interp}$^{\dagger}$ & Gamma-ray linear interpolation (0\,=\,off, 1\,=\,on; Sect.~\ref{sec:ni_ray_interp}) & 1 & 1 & \nodata \\
\hline
\multicolumn{5}{c}{\textit{Physics and EOS}} \\
\hline
\texttt{radiation} & Enable radiative transfer & 1 (on) & 1 (on) & \nodata \\
\texttt{eoskey} & Equation of state (1\,=\,ideal, 2\,=\,Paczy\'{n}ski) & 2 & 2 & \nodata \\
\texttt{saha\_ncomps} & Saha ionization species & 3 & 3 & \nodata \\
\texttt{boxcar\_smoothing} & Boxcar-smooth composition profile (pre-mapping) & 1 (on) & 1 (on) & \nodata \\
\texttt{boxcar\_smooth\_ni}$^{\dagger}$ & Enable Ni-specific boxcar smoothing & 1 (on) & 1 (on) & \nodata \\
\texttt{boxcar\_ni\_nominal\_mass\_msun}$^{\dagger}$ & Ni-specific boxcar mass window (Sect.~\ref{sec:ni_independent_smoothing}) & 0.4 & 0.4 & $M_{\odot}$ \\
\texttt{boxcar\_ni\_number\_iterations}$^{\dagger}$ & Ni-specific boxcar iterations (Sect.~\ref{sec:ni_independent_smoothing}) & 4 & 4 & \nodata \\
\texttt{smooth\_ni\_luminosity}$^{\dagger}$ & Photosphere Ni luminosity smoothing (Sect.~\ref{sec:ni_smoothing}) & 1 (on) & 1 (on) & \nodata \\
\texttt{opacity\_floor\_envelope} & Envelope opacity floor & 0.01 & 0.01 & cm$^{2}$\,g$^{-1}$ \\
\texttt{opacity\_floor\_core} & Core opacity floor & 0.24 & 0.24 & cm$^{2}$\,g$^{-1}$ \\
\hline
\multicolumn{5}{c}{\textit{Solver}} \\
\hline
\texttt{epstol\_hydro}$^{\dagger}$ & N--R convergence tolerance, hydro step & $10^{-7}$ & $10^{-4}$ & \nodata \\
\texttt{epstol\_rad}$^{\dagger}$ & N--R convergence tolerance, radiation step & $10^{-7}$ & $10^{-4}$ & \nodata \\
\texttt{itmax\_hydro}$^{\dagger}$ & Max N--R iterations, hydro step & 100 & 100 & \nodata \\
\texttt{itmax\_rad}$^{\dagger}$ & Max N--R iterations, radiation step & 300 & 300 & \nodata \\
\hline
\multicolumn{5}{c}{\textit{Timing and output}} \\
\hline
\texttt{tend} & Simulation end time & $5\times10^{6}$ & $5\times10^{6}$ & s \\
\texttt{dtmin} & Minimum allowed timestep & $10^{-10}$ & $10^{-10}$ & s \\
\texttt{dtmax} & Maximum allowed timestep $\Delta t_{\mathrm{max}}$ & $10^{4}$ & $10^{5}$ & s \\
\texttt{dtout} & Profile output cadence (late-time) & $1.7\times10^{5}$ & $5\times10^{5}$ & s \\
\texttt{dtout\_mid}$^{\dagger}$ & Profile output cadence (mid-time) & --- & $10^{5}$ & s \\
\texttt{dtout\_fast}$^{\dagger}$ & Profile output cadence (early-time) & --- & $5\times10^{4}$ & s \\
\texttt{dtout\_scalar} & Scalar output cadence (late-time) & $1.7\times10^{4}$ & $10^{5}$ & s \\
\texttt{dtout\_scalar\_mid}$^{\dagger}$ & Scalar output cadence (mid-time) & --- & $5\times10^{4}$ & s \\
\texttt{dtout\_scalar\_fast}$^{\dagger}$ & Scalar output cadence (early-time) & --- & $10^{4}$ & s \\
\texttt{output\_mid\_transition\_days}$^{\dagger}$ & Fast$\to$mid cadence transition & --- & 10 & d \\
\texttt{output\_late\_transition\_days}$^{\dagger}$ & Mid$\to$late cadence transition & --- & 20 & d \\
\texttt{output\_mode}$^{\dagger}$ & Output verbosity (0\,=\,full, 1\,=\,fitting-only; Sect.~\ref{sec:output_mode}) & 0 & 0 & \nodata \\
\enddata
\tablecomments{For original SNEC parameters, the ``Default'' column lists the SNEC~1.01 shipped value; for SuperSNEC-only parameters ($\dagger$), it lists the code default.  Solver tolerances show the SNEC~1.01 compiled constants for reference.  The \texttt{Ni\_period\_max} default is $10\times\texttt{Ni\_period}$ when not explicitly set.  SNEC~1.01 uses a single fixed output cadence; the three-band adaptive cadence is a SuperSNEC addition (Sect.~\ref{sec:cadence}).}
\end{deluxetable*}

\clearpage
\bibliographystyle{aasjournalv7}
\bibliography{adaptive_gridding_for_snec}

\end{document}